\documentclass[a4paper,onecolumn,11pt]{quantumarticle}
\pdfoutput=1
\usepackage[utf8]{inputenc}
\usepackage[english]{babel}
\usepackage[T1]{fontenc}
\usepackage{amsmath}
\usepackage{amssymb}
\usepackage{dsfont}
\usepackage{hyperref}

\usepackage{footnote}
\makesavenoteenv{tabular}
\usepackage{tikz}
\usepackage{lipsum}
\usepackage[numbers]{natbib}

\newcommand{\Hil}{\mathcal{H}}
\newcommand{\Tr}{\mathrm{Tr}}

\newcommand{\R}{\mathbb{R}}

\newcommand{\ket}[1]{|#1 \rangle}
\newcommand{\bra}[1]{\langle #1|}
\newcommand{\M}{\mathcal{M}}

\newcommand{\N}{\mathbb{N}}
\newcommand{\E}{\mathrm{E}}

\newcommand{\argmin}{\mathop{\mathrm{argmin}}} 
\newcommand{\half}{\frac{1}{2}}
\newcommand{\appn}{\mathrm{C}}

\def\figtext{Fig.~}

\newcommand{\ms}{\noalign{\vskip 0.5em}} 
\newcommand{\bhline}{\noalign{\hrule height 1pt}} 
\newcommand{\br}{\ms\bhline\ms} 
\newcommand{\mr}{\ms\hline\ms} 

\usepackage{xcolor}

\begin{document}

\title{QestOptPOVM: An iterative algorithm to find optimal measurements for quantum parameter estimation}

\author{Jianchao Zhang}
\email{c2141016@edu.cc.uec.ac.jp}
\affiliation{Graduate School of Informatics and Engineering, The University of Electro-Communications, Tokyo 182-8585, Japan}
\orcid{0009-0005-5299-6940}
\author{Jun Suzuki}
\email{junsuzuki@uec.ac.jp}
\homepage{http://www.q-phys.lab.uec.ac.jp/english/index.html}
\orcid{0000-0003-1975-6003}
\affiliation{Graduate School of Informatics and Engineering, The University of Electro-Communications, Tokyo 182-8585, Japan}
\maketitle

\begin{abstract}
Quantum parameter estimation holds significant promise for achieving high precision through the utilization of the most informative measurements. While various lower bounds have been developed to assess the best accuracy for estimates, they are not tight, nor provide a construction of the optimal measurement in general. Thus, determining the explicit forms of optimal measurements has been challenging due to the non-trivial optimization. In this study, we introduce an algorithm, termed QestOptPOVM, designed to directly identify optimal positive operator-valued measure (POVM) using the steepest descent method. Through rigorous testing on several examples for multiple copies of qubit states (up to six copies), we demonstrate the efficiency and accuracy of our proposed algorithm. Moreover, a comparative analysis between numerical results and established lower bounds serves to validate the tightness of the Nagaoka-Hayashi bound in finite-sample quantum metrology for our examples. Concurrently, our algorithm functions as a tool for elucidating the explicit forms of optimal POVMs, thereby enhancing our understanding of quantum parameter estimation methodologies.
\end{abstract}
\section{Introduction}

 Quantum metrology significantly enhances the precision of measurements on quantum probe states for various near-term applications, such as quantum phase estimation, quantum sensing, and quantum imaging \cite{kolobov2007quantum,giovannetti2011,dowling2015quantum,genovese2016real,degen17,pirandola2018advances,Albarelli2020, rodriguez2021efficient}. For instance, leveraging squeezed states of light enables the detection of weak signals in the laser interferometer gravitational wave observatory \cite{aasi2013enhanced}. The scientific study of measurements, known as quantum metrology, holds operational significance in finite-sample regimes, as realistic experiments are conducted with limited data sizes \cite{audenaert2012quantum,sugiyama2012adaptive,sugiyama2015precision,rubio2019quantum,meyer2023quantum}. This method revolves around estimating the underlying parameters of unknown quantum states, a concept initiated by Helstrom in the 1960s \cite{helstrom1967minimum,helstrom1968minimum,helstrom1969quantum,helstrom1976}. The mean squared error (MSE) serves as the relevant quantity for parameter estimation in statistics, with the objective being to minimize the MSE across all realizable measurements in laboratories. Consequently, the optimal measurement can extract the most information about the quantum states. Therefore, determining the extent to which the MSE can be reduced is a fundamental question from both theoretical and practical standpoints.

The ultimate precision of this value is determined by the quantum Cram\'er-Rao (CR) bound for single-parameter estimation. Helstrom introduced the symmetric logarithmic derivative (SLD) quantum Fisher information (QFI) for this bound, with subsequent studies revealing infinite families of quantum CR bounds \cite{petz1996monotone,petzbook}. However, these lower bounds based on the QFI are not always tight when estimating multiple parameters with an uncorrelated measurement strategy \cite{hayashi-book,sbd16,demkowicz2020multi,suzuki2020quantum}. Consequently, the ultimate precision for the MSE when estimating multiple parameters remains an open problem for finite-sample quantum metrology. Furthermore, obtaining an explicit form of the optimal measurements is crucial in practical applications.

To address this challenge, a numerical method can be employed to determine the minimum value of the MSE along with the mathematical expression for the optimal measurement, given by a positive operator-valued measure (POVM). The key idea underlying this study is that this value can be derived by optimizing the classical Fisher information matrix (CFIM) over all possible measurements \cite{young1975,nagaoka1989anew,braunstein1994statistical}. This approach provides all relevant information pertaining to the problem. Notably, Hayashi and Ouyang recently demonstrated that the ultimate bound can be formulated as conic programming as well \cite{hayashi2023tight}. Despite several attempts to implement these algorithms, previous methods fail to guarantee sufficient precision and/or computational efficiency as the size of the Hilbert space increases. For example, the recent Python library, QuanEstimation, does not perform full optimization over all possible measurements \cite{zhang2022quanestimation}, and conic programming has limitations in precision \cite{hayashi2023tight}. The algorithm proposed by Kimizu et al. only searches for rank-1 measurements \cite{kimizu2024adaptive}. Some of these drawbacks in existing algorithms could inherit from the use of solvers for optimization.

To overcome these limitations, we propose an efficient and accurate algorithm, specifically designed to determine the numerical value of the lowest MSE and the optimal measurement. We formulate the problem as a convex optimization problem and implement the steepest descent algorithm to iteratively optimize an objective function. We refer to this as QestOptPOVM, which is available as a MATLAB code \cite{QestOptPOVMgit}. We demonstrate that QestOptPOVM successfully identifies optimal measurements for estimating two parameters encoded in a six-qubit system (with a Hilbert space dimension of $2^6 = 64$) with a precision of the order of $10^{-6}$. QestOptPOVM not only efficiently identifies the optimal measurement numerically, but also enables the discovery of analytical forms by analyzing numerical results. This is feasible when a parametric family of quantum states exhibits a certain symmetry. To showcase the effectiveness of QestOptPOVM, we apply it to various problems in a qubit system and derive a family of optimal measurements with the smallest measurement outcomes. This provides insights into previously known optimal measurements with more outcomes.

The remainder of the paper is structured as follows: section 2 introduces the research, outlining the optimization problem setting and the lower bounds. Section 3 details the algorithm of QestOptPOVM, while sections 4 and 5 present two results addressing analytically solvable and numerically tractable models, respectively. Finally, in section 6, we conclude by summarizing all results and its implications.

\section{Preliminaries}

	\subsection{Problem setting}
	The problem at hand is to find a POVM for a given $n$-parameter model $ \{\rho_{\theta}|\theta\in\Theta \subset \R^n \}$ on the $d$-dimensional Hilbert space ${\cal H}=\mathds{C}^{d}$ where $\theta=(\theta_1,\dots,\theta_n)^T$ is the $n$-dimensional real vector. The partial derivative of the state with respect to $\theta_i $ is denoted as
    $\partial_i \rho_\theta=\frac{\partial \rho_{\theta}}{\partial \theta_{i}} \quad (i=1,2,\ldots,n$).
    The measurement is described by the POVM $\Pi=(\Pi_1,\Pi_2,\dots,\Pi_K) $, where $\Pi_k \geq 0$ and $\sum_{k=1}^K \Pi_k=I$ (the identity matrix on ${\cal H}$). The probability of receiving outcome $k$ is defined as $p_\theta(k|\Pi)=\Tr (\rho_\theta \Pi_k )$.
    
    The MSE is widely used in statistics to assess the accuracy of an estimator $\hat{\theta}$. In a multiparameter problem, it is defined as a MSE matrix
    \begin{align*}
        \left[ V_\theta(\Pi,\hat{\theta}) \right]_{i,j} =  \E_\theta[ (\hat{\theta_i}(X)-\theta_i  )(\hat{\theta_j}(X)-\theta_j  ) ] ,\ {\rm for }\ i,j=1,\dots,n ,
    \end{align*}
    where $X$ is the outcome, $\Pi$ is the measurement and $E_\theta[ \cdot ] $ is the expected value over the distribution of $p_\theta (k|\Pi)$. By defining the estimator for each outcome as $\hat{\theta}_i(X=k)=\hat{\theta}_{i,k}$, the $i,j$ component of MSE matrix becomes
    \begin{align*}
    \left[ V_\theta(\Pi,\hat{\theta}) \right]_{i,j} =  
     \sum_{k=1}^K (\hat{\theta}_{i,k}-\theta_i)(\hat{\theta}_{j,k}-\theta_j) \Tr (\rho_\theta \Pi_k )
      ,\ {\rm for }\ i,j=1,\dots,n.
    \end{align*}
    The goal is to minimize the MSE under the locally unbiased estimator condition,
    \begin{align}\label{locallyunbiased}
    \E_\theta [\hat{\theta}_i(X) ]=\theta_i \quad \mathrm{and} \quad  \frac{\partial}{\partial\theta_j} \E_\theta[ \hat{\theta}_i(X) ] = \delta_{i,j} ~\mathrm{at}~ \theta~\mathrm{for} ~ i,j=1,\ldots,n.    
    \end{align}
    It is widely known that the CR inequality holds for any locally unbiased estimator. 
    \begin{align*} V_\theta(\Pi, \hat{\theta}) \geq J_\theta (\Pi)^{-1}, 
    \end{align*}
    where $J_\theta (\Pi) $ is the CFIM :
    \begin{align*} [J_\theta (\Pi)]_{i,j}=\sum_{k=1}^K \frac{\Tr \partial_i \rho_\theta\Pi_k \Tr \rho_{\theta,j}\Pi_k }{\Tr \rho_{\theta}\Pi_k} .
    \end{align*}
    Note there always exists a locally unbiased estimator $\hat{\theta}$ for which the equality in this inequality holds at any local point as $\hat{\theta}_i (k)=\theta_i+\sum_j (J_\theta (\Pi)^{-1})_{i,j} \frac{\partial}{\partial \theta_j}(\log \Tr \rho_\theta \Pi_k).  $ Since there is no general definition of the minimal of a matrix, we prefer to use a certain function to derive a scaler from the MSE, which is known as the A-optimality in statistics \cite{fedorovbook,pukelsheimbook}. In this paper, we use the trace with a weight matrix $W>0 $:
    \begin{align*} \Tr (W V_\theta(\Pi, \hat{\theta})) \geq \Tr (W J_\theta (\Pi)^{-1}) .
    \end{align*}
    The weight matrix serves as a matrix to control the weight we assign to each parameter. For example, if we set $ W=I$, this indicates an equal importance across all parameters. Given the equality holding for any points, the objective shifts from minimizing the MSE to minimizing the weighted trace of the inverse of the CFIM. As we proceed with the problem setting, we are interested in finding an optimal POVM $\Pi$ at a given point $\theta$. We will assume that $\theta$ is given and, unless explicitly stated, we shall omit reference to the parameter $\theta$. For example, $\rho \equiv \rho_\theta, \partial_i \rho \equiv \partial_i \rho_\theta=\frac{\partial \rho_{\theta}}{\partial \theta_i} $ and $J(\Pi) \equiv J_\theta(\Pi)$.

\subsubsection{POVM optimization with weighted trace CFIM inverse}
~\\
 Let $\M_K $ be the set of POVMs on $\Hil$ whose outcomes are labelled by a set $\{1,\dots,K\}$, $K\in \N$
    \begin{align*}
    \M_K= \left\{  \Pi=(\Pi_1, \Pi_2,\dots,\Pi_K) ~  \left|~ \Pi_k \in \mathds{C}^{d \times d},  \Pi_k\geq 0, \Pi_k \neq 0,  \sum_{k=1}^{K}\Pi_k=I  \right\}\right..
    \end{align*}
    In this set, we emphasize that the zero measure that has no contribution is discarded. The size of a POVM, also means the number of elements in a POVM, is denoted as $K$.
    
    Then, we formulate the first problem as follows. Given the state $\rho$ and its derivatives $\partial_j \rho$ ($j=1,2,\ldots,n$),
	find an optimal POVM with a given size $K$ such that it minimizes the trace of the inverse of the CFIM, 
    \begin{align*}
		\min_{\Pi \in \M_K } \Tr \left( W J(\Pi) ^{-1} \right) ,
	\end{align*}
    whose $i,j$ component is
    \begin{align*} [J(\Pi)]_{i,j}=\sum_{k=1}^K \frac{\Tr \partial_i \rho \Pi_k \Tr \partial_j \rho \Pi_k }{\Tr \rho \Pi_k} .
    \end{align*}
    The optimal POVM, which gives the minimal value and has $K$ outcomes is donated as $\Pi^*_K$,
    \begin{align*}
    \Pi^*_K=\argmin_{\Pi \in  \M_K } \Tr \left( W J(\Pi) ^{-1} \right).
    \end{align*}
    Note that $\Pi_K^*$ is not unique in general. Examples of non-uniqueness of optimal POVMs are discussed in section \ref{result1}.

	\subsubsection{Minimum outcomes of optimal measurements}
    ~\\
    It is known that this problem can be formulated as a convex optimization problem since $\M_K$ is a convex set and $\Tr (J(\Pi)^{-1})$ is a convex optimization problem \cite{fujiwara06consistency}. Due to Carath\'eodory's theorem, the size of POVM has a certain upper bound \cite{kimizu2024adaptive} as $\frac{1}{2}d(d+1)+n(n+1)$.
    Let the set of all optimal POVMs as
    \begin{align*}
    \M^* = \bigcup_{K \in \mathbb{N}} \M_K^*,    \quad \mathrm{where}\ 
    \M_K^*=\bigg\{ \Pi^*_K=\argmin_{\Pi \in  \M_K } \Tr \left( W J(\Pi) ^{-1} \right)   \bigg\}.
    \end{align*}
    Then we are interested in the minimal size of optimal POVMs which is denoted as $K^*$. Here the $K^*$ means the minimal number of outcomes for all possible optimal POVMs. Mathematically, this is denoted as
    \begin{align*}
    K^*=\min_{\Pi^*\in \M^* } \left\{ K ~|~ \Pi^* \mathrm{\ has \ } K \mathrm{\ outcomes} \right\}   .
    \end{align*}
    If we define the set of POVMs which is both optimal and has minimal size, $\M_{\min} =\{ \Pi^* | \Pi^* \in \M_K^*,~ \Pi^* \mathrm{\ has \ } K^* \mathrm{\ outcomes} \}$, this set is a boundary of all POVMs. 
    
There is a natural convex structure on $\M^*$ by randomized combination, which is useful in parameter estimation \cite{fujiwara06consistency,yamagata2011,suzuki2021quantum}. In passing, another convex structure on $\M_K$, which is defined by elementwise addition, is also used in quantum information theory \cite{d2005classical}. However, there is no trivial convex structure on $\M_{\min}$. In the later section \ref{result1}, we aim to find the analytical form of $\M_{\min}$ for certain solvable cases.
	
 \subsection{Necessary and sufficient condition for qubit models}
    The elementary example of estimating qubit parameters can be solved analytically \cite{nagaoka1989anew,hayashi97,GM00}. In previous studies, this was solved with randomized PVM, see a unified approach for the optimal POVM \cite{yamagata2011}. A necessary and sufficient condition of optimality and a construction of randomized POVM are given by using the SLD \cite{yamagata2011}. For a given weight $W$
    \begin{align*}
    \min_{\Pi:\mathrm{POVM}} \{\Tr \left(W J(\Pi)^{-1}\right) \}= (\Tr R)^2 ,
    \end{align*}
    where $ R=\sqrt{J_S^{-\frac{1}{2}} W J_S^{-\frac{1}{2}}  }$ with $J_S$ the SLD QFI matrix. The minimum is attained if and only if $\Pi  $ satisfies 
    \begin{align}\label{necandsuf}
        J(\Pi) = \frac{ \sqrt{J_S}R \sqrt{J_S}}{\Tr R} .     
    \end{align}
    The optimal measurement is the random combination of a normalized operator. Diagonalize $R$ as $R=U \Lambda U^{-1}$ where $\Lambda=\mathrm{diag}(\lambda_1,\dots,\lambda_n)$. The proportion of each eigenvalue is $p_i=\frac{\lambda_i}{\lambda_1+\dots+\lambda_n}$.
    Let $L_i$ be the $i$-th SLD defined by the equation,
    \begin{align*}
    \partial_i \rho = \frac{1}{2}(\rho L_i +L_i \rho).
    \end{align*}
    The SLD QFI matrix is written in
    \begin{align*}
    [J_S]_{i,j} = \frac{1}{2} \Tr [ \rho (L_i L_j +L_j L_i)].
    \end{align*}
    Define a Hermitian matrix, 
    \begin{align*}
    L^i=\sum_{k=1}^n  \left( U^{-1} \sqrt{J_S^{-1}} \right)^{i,k} L_k.
    \end{align*}
    Let $\Pi^{(i)} $ be a projection-valued measure (PVM) given by the spectral decomposition of $L^i$. The optimal measurement is the random combination of them, which is called randomized measurement. This is realized by taking measurement $\Pi^{(i)}$ with probability $p_i$ and written as
    \begin{align*}
    \Pi^* \equiv p_1 \Pi^{(1)} \oplus \dots \oplus p_n \Pi^{(n)}.
    \end{align*}
     For example, given two PVMs $\Pi^{(1)},\Pi^{(2)}$; $\Pi_\pm^{(1)}$ measures $\sigma_1$ and $\Pi_\pm^{(2)}$ measures $\sigma_2$.
    Then the randomized measurement is
    \begin{align*}
    \Pi= \left( p_1 \Pi_-^{(1)} ,p_1 \Pi_+^{(1)} ,p_2 \Pi_-^{(2)} ,p_2 \Pi_+^{(2)}\right),
    \end{align*}
    where $p_1$ and $ p_2$ are corresponding probability of measuring each PVM.
    
    Thus, by using this measurement, the number of outcomes$=\dim \mathcal{H} \cdot n$. Indeed, in the qubit case $(\dim \Hil =2)$, this size $2n$ is not the minimal size $K^*$. Finding $K^*$ of qubit case is explored in this paper.

	\subsection{Lower bounds} \label{lowerbound}
    This paper intends to provide an accurate and efficient algorithm to compute the optimal value together with corresponding POVMs. Since this is a numerical result, assessing it by the speed of convergence is not sufficient to announce its reliability. A more convincing method is to compare the solution with a lower bound of the optimal question. If there is a numerically neglectable difference between the lower bound and minimized quantity, this makes the lower bound saturated and infers that the measurement found by our algorithm is one of the optimal choices. 

    The widely known lower bound, the Holevo bound, is a general one but it requires infinitely large collective measurement to achieve asymptotically \cite{holevobook}. A better or tighter alternative is the Nagaoka bound which is valid for any two-parameter model. It is known that it can be saturated in two-dimensional Hilbert space and the measurements are restricted to single-copy measurements \cite{nagaoka91}. Recently, this bound has been extended into the Nagaoka-Hayashi (NH) bound, capable of accommodating more than two parameters. It stands as the tightest known bound and it is computational tractability through semidefinite programming (SDP) \cite{conlon2021efficient}. Utilizing an efficient SDP solver facilitates rapid and accurate minimization of the duality gap. The NH bound degenerates to the Nagaoka bound in the context of a two-parameter model.
 
    To introduce the bounds, it is convenient to define a set of unbiased operators by
    \begin{align*}
    X_j \equiv \sum_k \hat{\theta}_{j,k}\Pi_k - \theta_j I .
    \end{align*}
    Thus the locally unbiased condition Eq.~(\ref{locallyunbiased}) is expressed as \cite{nagaoka1989anew}
    \begin{align}\label{luo}
    \Tr \rho X_j =0~ ~\mathrm{and} ~~ \Tr \partial_k \rho X_j =\delta_{j,k} \quad \mathrm{for}\ j,k=1,\ldots, n  .
    \end{align}
    
    The weight matrix signifies the weighting assigned to the significance of each parameter. In subsequent discussions, we adopt the identity matrix, denoted as $W=I$, as the weight matrix to streamline the analysis. This choice implies equal significance is attributed to each parameter. Importantly, this assumption does not compromise the generality of our approach, as it can be extended to encompass any weight matrices \cite{fn99}.
    \subsubsection{Holevo bound}
    ~\\
    The Holevo bound can be written as \cite{nagaoka1989anew,holevobook}
    \begin{align*}
    c_\mathrm{H} \equiv \min_{X=(X_1,X_2,\ldots,X_n)}
    \big\{   \Tr  \mathrm{Re} Z(X) +\Tr | \mathrm{Im} Z(X) | \,\big|\, X_j: \mathrm{Hermitian} \ \mathrm{satisfying} \ (\ref{luo})\big\},  \end{align*}
   where $Z(X)$ is an $n \times n$ Hermitian matrix with $Z_{j,k}(X)=\Tr (\rho X_kX_j)$. and Hermitian $X_j$ satisfies Eq.~(\ref{luo}). $\mathrm{Re}(Z)$ means the real part of $Z$ and $\mathrm{Im}(Z)$ means the imaginary part of $Z$. $|X|=\sqrt{X^\dag X}$. In other words, $\Tr|X|$ is the absolute sum of all eigenvalues of $X$.
    \subsubsection{Nagaoka bound}
    ~\\
	The Nagaoka bound is a lower bound for the trace of MSE.
	\begin{align*} c_\mathrm{N} \equiv \min_{X_1,X_2} \big\{ \Tr \left[ \rho X_1 X_1 + \rho X_2 X_2 \right]
	+\Tr | \sqrt{\rho} [X_1,X_2] \sqrt{\rho} | \,\big|\, X_1,X_2: \mathrm{Hermitian} \ \mathrm{satisfying} \ (\ref{luo})\  \big\} .
	\end{align*}
	
	\subsubsection{Nagaoka-Hayashi bound}
	~\\
    The NH bound is the extension of the Nagaoka bound for more than two parameters.
	\begin{align*} c_{\mathrm{NH}} \equiv \min_{\mathbb{L},X} \big\{ \mathbb{T}{\rm r} [ \mathbb{S}_\theta \mathbb{L} ]\ \big| \ \mathbb{L}_{j,k}=\mathbb{L}_{k,j}: {\rm Hermitian}, \mathbb{L}\geq XX^T, X_j:\ \mathrm{Hermitian} \ \mathrm{satisfying} \ (\ref{luo})\ \big\} .
	\end{align*}
  where $\mathbb{S}=I_n \otimes \rho  $ and $ \mathbb{L} $ is an $n\times n$ block matrix value of Hermitian matrices, which means $\mathbb{L} \in \mathcal{L}( \mathbb{C}^n \otimes \Hil)=\mathbb{C}^{nd \times nd} $. There are two symbols of trace. 
    \begin{itemize}
        \item $\mathbb{T}\mathrm{r} [\cdot]$ means the trace over both parametric space and quantum systems $\mathbb{C}^n \otimes \Hil$.
        \item $\Tr[\cdot] $ means the partial trace over the quantum system $\Hil$.
    \end{itemize} 
   To compute this bound is equivalent to solving an SDP problem which can be written as \cite{conlon2021efficient}:
    \begin{align*}
        c_{\mathrm{NH}}=\min_{\mathbb{L},X}\mathbb{T}\mathrm{r}[\mathbb{S}_\theta \mathbb{L}],\\
        {\rm subject\ to} \begin{pmatrix}
            \mathbb{L} & X \\
            X^T & 1  
        \end{pmatrix}\geq 0 ,
    \end{align*}
    where $ \mathbb{L}_{j,k}=\mathbb{L}_{k,j},$ for all $j,k=1,\ldots,n$. $\mathbb{L}_{j,k}$ and $X_j$ are Hermitian and $X_j$ satisfying Eq.~\eqref{luo}.
     As mentioned before, the Nagaoka bound is compatible with the two-parameters model. Thus, we compare the minimum value found by this research with the Nagaoka bound in this model and the NH bound in more than two parameters model. The Holevo bound is only asymptotically achievable which means there is an unavoidable gap between the Holevo bound and the optimal value for a finite number of collective measurements \cite{conlon2022gap}. It is a well-established fact that by definition the NH bound is tighter than the Holevo bound, $c_\mathrm{NH} \geq c_\mathrm{H}$.
    It is acknowledged that the NH bound degenerates in the Nagaoka bound in the two-parameters model. The NH bound reduces to the Gill-Masser bound in the qubit model. In the following, we shall use the phrase `the NH bound' when referring to the Nagaoka bound for two-parameter estimation and the Gill-Masser bound for qubit models for simplicity.
    
	\subsection{Related works}
	The QuanEstimation is a toolkit for quantum parameter estimation \cite{zhang2022quanestimation}. It has various optimization methods including control optimization, state optimization, and measurement optimization. Measurement optimization is the part we are interested in. Briefly, QuanEstimation can find measurements in these three different cases. 
    \begin{enumerate}    
        \item Rank-one projective measurements.
        \item Linear combination of a given set of POVMs. 
        \item Optimal rotated measurement of an input measurement.
    \end{enumerate}
    It is clear that these options do not exhaust all possible measurements. 
    
    There is another numerical method to find the optimal POVM by parameterizing the rank-one measurement into the two-level orthogonal matrix \cite{kimizu2024adaptive}. To optimize over $K$-valued  rank-one measurement, the completeness relation is written in 
    \begin{align*}
    \sum_{k=1}^K \ket{a_k} \bra{a_k}=I 
    \Leftrightarrow
    \begin{pmatrix}
        \ket{a_1} & \ket{a_2} & \ldots & \ket{a_K} 
    \end{pmatrix}
    \begin{pmatrix}
        \bra{a_1} \\ \bra{a_2} \\ \vdots \\ \bra{a_K} 
    \end{pmatrix}=I .
    \end{align*}
    If we let $V^\dag= \begin{pmatrix}
        \ket{a_1} & \ket{a_2} & \ldots & \ket{a_K} 
    \end{pmatrix} $, then $V$ is an isometry. This means that there exists $m$ two-level orthogonal matrices $U_1,U_2,\ldots,U_m \in \R^{K\times K}$ such that 
    \begin{align*}
    U_1 U_2 \ldots U_m V = \begin{pmatrix}
        1 & 0 & \ldots & 0 \\
        0 & 1 & \ldots & 0 \\
        \vdots & \vdots & \ddots & \vdots \\
        0 & 0 & \ldots & 1 \\
        0 & 0 & \ldots & 0 \\
        \vdots & \vdots & \vdots & \vdots \\
        0 & 0 & \ldots & 0
    \end{pmatrix} = \begin{pmatrix}
        I \\ 0
    \end{pmatrix} \in \R^{K \times d}.
    \end{align*}
    The number $m$ equals to $(K-1)+(K-2)+\ldots+(K-d)=Kd-\frac{1}{2}d(d+1)$. Optimizing $V$ is equivalent to finding each $U_i$ which can be specified with one single parameter. This means that minimizing the rank-one measurement becomes optimizing over $m$ parameters.

    As a complement to these measurement optimizations, our algorithm starts with a more general setting of measurement. The rank of measurement is not constrained in the process of optimizing. Abstractly, since the set of measurements $\M $ is a convex set, going through the road of rank-one POVMs means finding the optimal one in a subset, the subset contains only rank-one POVMs. Releasing the control of rank one is a more efficient choice. Moreover, this helps us to find the $K^*$ which is the core of defining the extreme points of optimal measurements.


\section{Algorithm of QestOptPOVM}
    The proposed algorithm is based on the steepest descent method. Instead of working directly on POVMs, we introduce the Kraus operator (the measurement operator) $A=(A_1, A_2 \ldots, A_K),$ where $A_k \in \mathbb{C}^{d \times d} $. The relation between this operator and POVM is $A_k^\dag A_k = \Pi_k$. This guarantees the positivity condition since $\Pi_k=A_k^\dag A_k \geq 0$. Our objective function is $ \Tr (J( \Pi )^{-1})$. We define a new function with respect to $A$ and use the same symbol $J$ for convenience as $ \Tr (J( A )^{-1})$. The practical advantage of using the Kraus operator is to avoid the situation when we update the operator, it is possible that the measurements become non-positive because of some tiny non-zero value.

    The other condition we need to keep is the completeness relation $\sum_k \Pi_k=I $. This can be formulated in the Lagrange multiplier method \cite{boyd2004convex}. Introduce $\Lambda \in \mathbb{C}^{d \times d}$, then applying this strategy we add a constraint in the objective function and have one more matrix $\Lambda$ to optimize. The objective function we need to minimize is 
    \begin{align*}
    f(A,\Lambda)=  \Tr (J(A)^{-1})+  \Tr \left(\Lambda \Big( \sum_{k=1}^K A^\dag_k  A_k -I \Big) \right).
    \end{align*}
    The next step of the gradient descent method is to compute the first-order derivative of this objective function. The detail of this calculation is written in Appendix A. This then gives an update as $A_k^{\mathrm{(new)}}=A_k^{\mathrm{(old)}}+\alpha H_k $ such that the change of the objective function remains non-positive. This relation is under two specifies. The measure is the Hilbert-Schmidt inner product and the update step size $\alpha$ is sufficiently small.

    The last step is to compute the Lagrange multiplier $\Lambda^{(m)}$ in this case and obtain the iteration equation of this algorithm. We set $H_k^{(m)}$ is the update matrix of $m$-th iteration and $\alpha$ is the step size.  The $m$-th iteration would be:
    \begin{align*}
        A^{(m+1)}_k &= A^{(m)}_k+\alpha H^{(m)}_k\\
        &=  A^{(m)}_k\left( I+ \alpha (  X^{(m)}_k - \Lambda^{(m)}) \right),
    \end{align*}
    where $X^{(m)}_k $ and $ A_k^{(m)}$ are given by equations (\ref{Akn}) and (\ref{Xkn}) in Appendix A, respectively.

    Even though we carefully choose the step size $\alpha$, note $X_k^{(m)},\Lambda^{(m)}$ depend on $\rho,\partial_i \rho, A_k^{(m)}$, the completeness relation still cannot be satisfied exactly after iterations. This is simply because our method is provided valid only up to the first order in $\alpha$. To improve the accuracy of the algorithm, we use a renormalization method after each update. This method is to multiply the square root inverse of the sum of all measurements on each side of the measurement. This step will be explained in section \ref{normalization}. 

    As a gradient descent method, the stopping rule is required to be set. The common stopping rule is to control the maximal number of iterations $m_{\max}$. In this paper, we additionally consider the variation of each step. The detail is written in section \ref{linesearch}.

    The greatest advantage of this algorithm compared with other programming optimization algorithms is that we take the advantage of keeping the form of the matrix in calculating the derivative. Other gradient methods based on changing each input item and regarding the input as a long vector will definitely not be as efficient as this algorithm. Compared to simple vector optimization with a nonlinear function, the qubit case with four POVM elements will need $4*(2^2)=16$ real numbers with four positivity inequalities and one completeness relation equation. Another advantage of our algorithm lies in its applicability to both rank-deficient models and pure-state models. This is because our algorithm relies only on $\rho,\partial_i \rho$, but not on $\rho^{-1}$.

\subsection{Performance of the proposed algorithm}
\begin{figure}[htbp]
    \centering
    \includegraphics[width=8cm,page=1]{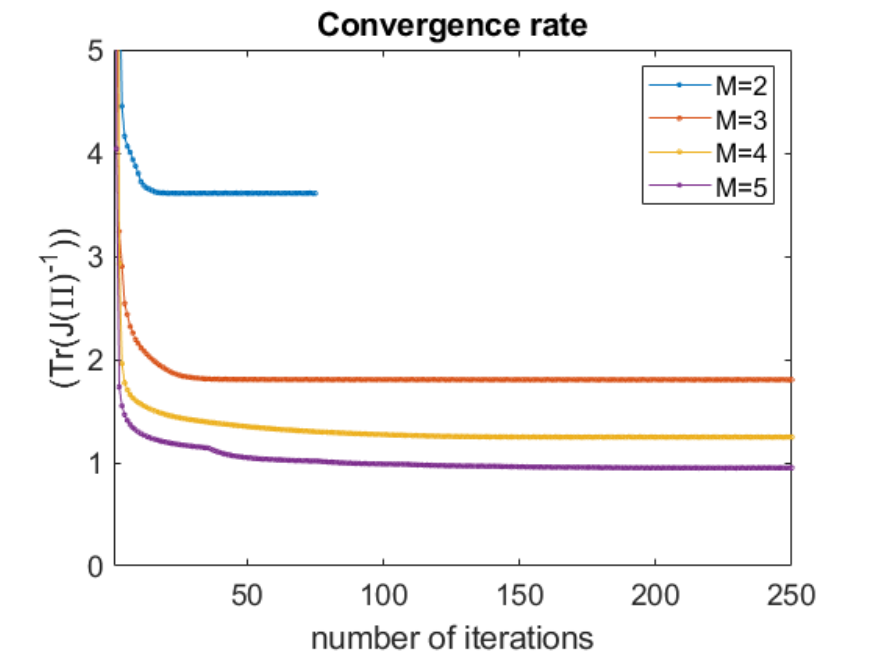}
    \caption{The objective function $\Tr [J(\Pi)^{-1}]$ versus the number of iterations from two to five copies shows the performance of the algorithm. The model is the three-parameter qubit with multiple copies model. See the details of the model in section \ref{qubitthreeparameters}.  } \label{Fig:perf}
\end{figure}

The convergence rate of multiple copies ranging from two to five qubits is depicted in \figtext \ref{Fig:perf}. In this context, ``$M$ qubits" denotes the simultaneous estimation of $M$ copies of a qubit by collective POVMs on them. Notably, the case of two copies converges to its limit at approximately seventy iterations due to the stopping rule. From a practical standpoint, an optimal value is achieved very closely after merely ten iterations. However, for three or more copies, the endpoint does not occur within 250 iterations. The convergence rate curve for three qubits appears relatively flatter compared to that of four and five qubits. The computational time scale is listed in table \ref{tabletime}, revealing an exponential increase in computational cost with the number of copies. However, this is a reasonable increase rate corresponding to the increased matrix size.

\begin{table}[ht] 
\centering
\begin{tabular}{l | l| l | l | l}
\hline
Number of copies $M$ & 2 & 3 & 4 & 5 \\ \hline
Time cost (seconds) & $< $0.1 & $\approx$1 & $\approx$10 & $\approx$100 \\ \hline
\end{tabular}
\caption{The approximate computing time versus the number of copies shows the exponential relation between these two values. Computer configuration: Processor Intel(R) Core(TM) i7-10700 CPU @ 2.90GHz, memory 16GB. Matlab version: 2022a.} \label{tabletime}
\end{table}

\subsection{Line search and stopping rule} \label{linesearch}
Line search is an approach to determine the value of step size \cite{boyd2004convex}. In QestOptPOVM, we use a naive way that calculates changes for each alternative which is pre-chosen and then chooses the one that has minimal value in each step. 
The stopping rule is the strategy to constrain the time of iteration. This algorithm sets two stopping rules, one is checking if the variation of each step is smaller than a very tiny value, and the other one is the maximal number of iterations. 
If the variation $\epsilon$, which is defined by the change in each iteration, is small enough (namely $10^{-10}$ by default in this algorithm) the algorithm will stop immediately at the current iteration. This tiny difference does not mean it stops at the optimal result but it means that the remaining iterations do not affect a lot. Besides, the number of maximal iterations, 1000 as default, is set to control the time. 

\subsection{Renormalization to keep the completeness relation} \label{normalization}
In the algorithmic context, the utilization of the Lagrange multiplier method guarantees the completeness relationship, contingent upon employing a sufficiently diminutive step size, which proves impracticable in optimization scenarios. In practice, a medium step size is used, leading to the violation of the completeness relationship. To address this issue, a recalibration of the POVM is executed subsequent to each iteration to keep the completeness relation \cite{somim}. This process precludes the distortion of the POVM and enhances the accuracy of the algorithm. Specifically, the recalibration entails multiplying both sides of the equation by the square root of the inverse of the summation. Let $G = \sum_k A_k^\dag A_k >0$, then define $\bar{A_k} =A_k G^{-\frac{1}{2}}$. Consequently, the completeness relation is ensured: $\sum_k \bar{A_k}^{\dag} \bar{A_k}=I$. It is worth noting that this recalibration process necessitates the computation of the exact inverse of the measurement summation, thereby contributing to the computational complexity, particularly in scenarios involving tensorial states.

	
\section{Result1: analytically solvable models}\label{result1}
 
   Our algorithm, QestOptPOVM provides us an opportunity to numerically find the optimal POVMs for any fixed parameter $\theta$ and size of POVM, $K$. However, in almost all models, the optimal POVM is not unique. In order to find the set of all optimal POVMs, what we are going to do is to get different optimal POVMs by feeding different seeds for initialization. To confirm the POVMs are optimal, we compute the corresponding NH bound by SDP \cite{conlon2021efficient,scs}. We will demonstrate that it is feasible to construct the general form of optimal POVMs by looking at these different optimal POVMs in a few parameter cases. The cases in which we are capable of finding the analytical form of optimal POVMs are written in this section. The first case in this section is straightforward. 

    Since this algorithm can find not only the optimal POVM but also the minimal number of POVM elements $K^*$, in this section the optimal POVM with the minimal number of outcomes is discussed. The POVM with more elements is not included.
 
    \subsection{Optimal POVM with symmetry}

        \subsubsection{Two-parameter qubit} \label{qubittwoparameters}
    ~\\
    Since we are solving the multiparameter problem, a two-parameter model is the first setting. Consider a two-parameter model $\rho_\theta=\frac{1}{2}(I+\theta_1 \sigma_1 +\theta_2 \sigma_2)$ where $\sigma_1,\sigma_2 $ are the Pauli matrices. We aim at finding optimal POVMs at $\theta=(\theta_1,\theta_2)=(0,0)$.

\begin{figure}[htbp]
    \centering
    \includegraphics[width=8cm,page=2]{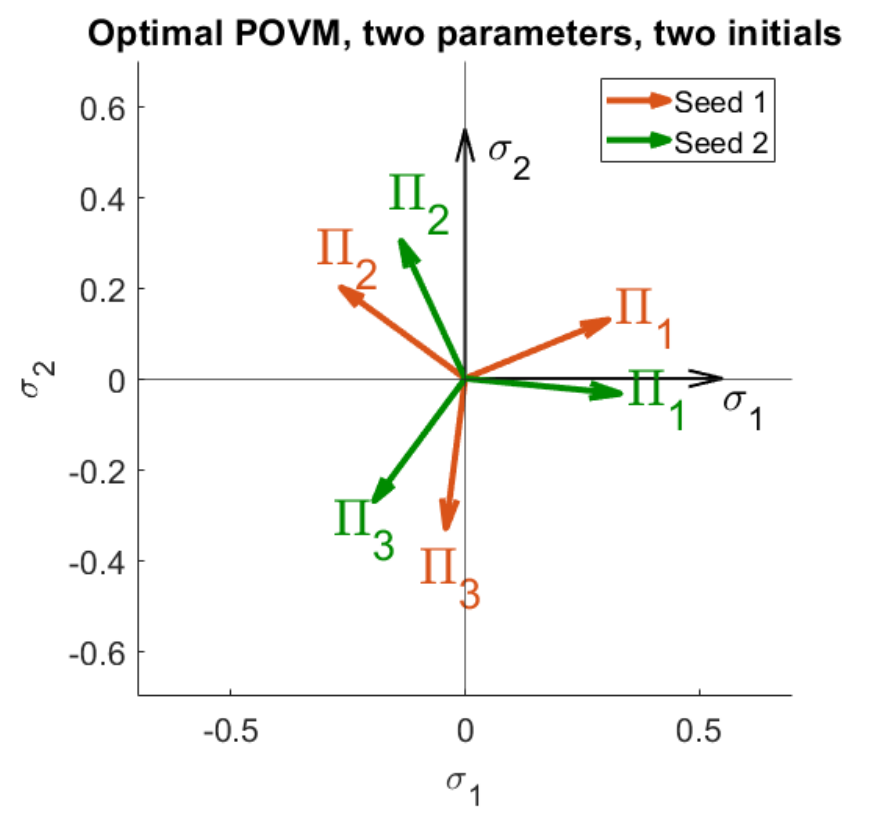}
    \caption{The coefficients of the optimal POVM on the directions of $\sigma_1$ and $\sigma_2$ when the parameters at origin form in three arrows. Two different initial seeds give two colors of arrows. Each color of the arrows forms an equilateral triangle. The model is the two-parameter Bloch model.  } \label{figorigin}
\end{figure}

    \begin{center}
    \begin{tabular}{@{}llll}
    \br
     state $\rho$   & derivatives $\partial_i \rho $ &  $ K^*$  & NH bound\footnote[1]{ This numerical value coincides with the analytical result.} \\
    \mr
     $\rho=\frac{1}{2} I $   &   $ \partial_1 \rho =\frac{1}{2} \sigma_1 $, $\partial_2 \rho =\frac{1}{2} \sigma_2 $   &  3  & 4.000 \\
    \br
    \end{tabular}    
    \end{center}
The optimal POVM was stated previously but did not refer to the minimal elements, see for example \cite{yamagata2011}. Finding the analytical form of optimal POVM in this example is possible from numerics. The diagonal components of POVM keep the same value. And the off-diagonal variate in a symmetric relation. By checking the magnitude we can ensure the off-diagonal entries are complex numbers on the unit circle with different angles. After we plot these complex numbers in \figtext \ref{figorigin} we noticed that the angle of these three measurements forms an equilateral triangle. 

    Based on these observations, we make an ansatz that the POVMs is the form: \\
    There exist three angles $\phi_1,\phi_2,\phi_3 $ such that
	\begin{align*}
		\Pi_k=\frac{1}{3} \begin{pmatrix}
			1 & e^{i\phi_k} \\
			e^{-i\phi_k} & 1
		\end{pmatrix},\ k=1,2,3,
	\end{align*}
	with \begin{align*} \phi_2=\phi_1+\frac{2}{3}\pi, \phi_3=\phi_1-\frac{2}{3}\pi. \end{align*} 
	In other words, $\{ \phi_1,\phi_2,\phi_3 \}$ is a set of three angles that forms an equilateral triangle. It is straightforward to show this POVM indeed satisfies the necessary and sufficient condition Eq.~\eqref{necandsuf}. Thereby, we prove the optimality of the POVM with the trine structure. This means that each rotation angle corresponds to one optimal POVM.

    In contrast, the optimal POVM of the same setting was given in the randomized measurement in previous studies. Consider two PVMs,
    \begin{align*}
    \Pi_\pm^{(1)}=\frac{1}{2} (I\pm \left(  \sin \phi ~ \sigma_1 - \cos \phi~ \sigma_2  \right) ),~\Pi_\pm^{(2)}=\frac{1}{2}(I\pm \left( \cos \phi ~ \sigma_1 + \sin \phi ~ \sigma_2 \right) ),
    \end{align*}
    and define the POVM with equal probability
    \begin{align*}
    \Pi=  \left( \frac{1}{2}\Pi_-^{(1)} , \frac{1}{2}\Pi_+^{(1)} , \frac{1}{2}\Pi_-^{(2)} , \frac{1}{2}\Pi_+^{(2)}\right),
    \end{align*}
    with $\phi$ a free parameter $\in [0,2\pi ) $.

    We noticed that the optimal POVM found previously exhibits coefficients exclusively in the direction of $\sigma_1$ and $\sigma_2$, corresponding to the parameters of the considered state. If we change the model by a two-parameter state with another pair of Pauli matrices, the optimal POVM similarly aligns along the same axes. This inherent symmetry suggests a parameterized model for the optimal POVM in the context of qubits with three parameters which is elaborated upon in the next section. 

    In this section, we yield the analytical result by fixing the local position of $\theta$ at a particular point, namely the origin. This choice affords the possibility of establishing symmetric relations across all directions. However, in section \ref{nontrivial}, we deviate from this constraint, allowing $\theta$ to take on any feasible value. While the analytical expression can still be derived, it becomes more intricate in form.

    \subsubsection{Three-parameter qubit}
	~\\
    The three-parameter qubit case is an extension of two parameters and it is the complete parametric space of a qubit. The model is $\rho_\theta=\frac{1}{2}(I+\theta_1\sigma_1 +\theta_2\sigma_2 +\theta_3 \sigma_3)$, and we set $\theta=(0,0,0).$
    \begin{center}
    \begin{tabular}{@{}llll}
\br
state $\rho$   & derivatives $\partial_i \rho $  & $K^*$  & NH bound\footnote[1]{ This numerical value coincides with the analytical result.}
\\
\mr
$\frac{1}{2} I  $   &   $\partial_1 \rho =\frac{1}{2} \sigma_1 $, $\partial_2 \rho =\frac{1}{2} \sigma_2 $ , $\partial_3 \rho = \frac{1}{2}\sigma_3 $  \quad \quad  &  4  & 9.000 \\
\br
\end{tabular}    
    \end{center}
In this case, the optimal measurement derived in a randomized measurement has six outcomes. In contrast, our algorithm, QestOptPOVM numerically finds optimal POVMs with four outcomes. We notice that the optimal solution is not unique. Then we parametrize the solution with a point in the Bloch sphere as
\begin{align*}
    \Pi_k=\frac{1}{4} \left( I+a_k \sigma_1+b_k \sigma_2+c_k \sigma_3 \right), k=1,2,3,4 .
\end{align*}
By plotting these four points $(a_k,b_k,c_k), k=1,2,3,4 $ in three-dimensional space, we find that all the non-unique results form a regular tetrahedron but different angles. This is what we expected from the previous section. There are three free parameters for choosing a regular tetrahedron. We provide the proof of optimality of this analytical form in Appendix B.
We remark that this tetrahedron measurement is known to provide minimal tomography \cite{rehacek2004minimal}.
\subsubsection{Two-parameter, two copies of qubit}
\label{qubittwoparatwocopy}
~\\
It is known that a collective measurement on two independent and identically distributed (i.i.d.) states can increase the MSE of estimation. The state of two copies can be written as the tensor product of two states, $\rho \otimes \rho$. In this paper, $M$ is the notation for the number of copies. In other words, $\rho^{\otimes M}=\underbrace{\rho \otimes \rho \otimes \ldots \otimes \rho \otimes \rho}_M $. Consider the same two-parameter qubit model as in section \ref{qubittwoparameters}.
\begin{center}
    \begin{tabular}{@{}llll}
    \br
     state $\rho$   & derivatives $ \partial_i \rho$  & $ K^*$  & NH bound \\
    \mr
     $ \rho^{\otimes 2} = \rho \otimes \rho $ with $ \rho=  \frac{1}{2}  I  $  &   $ \partial_1 \rho = \half \sigma_1 \otimes \rho +\rho \otimes \half \sigma_1 $, 
      &  4  & 1.500 \\
        &  $ \partial_2 \rho =\half \sigma_2 \otimes \rho + \rho \otimes \half \sigma_2 $  &   &  \\
    \br
    \end{tabular}
\end{center}
    We could notice that the Nagaoka bound in this setting is less than half of it in section 4.1.1 which is 4. This is one advantage of using i.i.d. state. The MSE is less than estimating one state multiple times.

    The minimal number of optimal POVM elements is four. One of them remains the same as $\Pi_4$ in the computational basis,
	\begin{align*} 	\Pi_4=\begin{pmatrix}
		0 & 0 & 0 & 0 \\
		0 & \frac{1}{2} & -\frac{1}{2} & 0 \\
		0 & -\frac{1}{2} & \frac{1}{2} & 0 \\
		0 & 0 & 0 & 0
	\end{pmatrix}=\ket{\psi^-}\bra{\psi^-},
	\end{align*}
	where $\ket{\psi^-}=\frac{1}{\sqrt{2}}(\ket{01}-\ket{10})$ is the singlet state. Note that $\Pi_4$ gives zero information about the parameters.
	
	The other three POVMs have constants on diagonal position. Other entries have the same magnitude and different angles. Based on our knowledge of the two-parameter problem, it is natural to plot the angles on the plane and find the symmetric relation. 
	\begin{align*}
		\Pi_k=& \begin{pmatrix}
			\frac{1}{3} & \frac{\sqrt{2}}{6}e^{i\varphi_k} & \frac{\sqrt{2}}{6}e^{i\varphi_k} & \frac{1}{3}e^{i\phi_k}\\
\frac{\sqrt{2}}{6}e^{-i\varphi_k} & \frac{1}{6} & \frac{1}{6} & \frac{\sqrt{2}}{6}e^{i\varphi_k} \\
\frac{\sqrt{2}}{6}e^{-i\varphi_k} & \frac{1}{6} & \frac{1}{6} & \frac{\sqrt{2}}{6}e^{i\varphi_k} \\
\frac{1}{3}e^{-i\phi_k} & \frac{\sqrt{2}}{6}e^{-i\varphi_k} & \frac{\sqrt{2}}{6}e^{-i\varphi_k} & \frac{1}{3} 
		\end{pmatrix} \\
		=&\ket{\psi_k}\bra{\psi_k}\ ({\rm rank~one}),
	\end{align*}
	where
	$$ \ket{\psi_k}=\begin{pmatrix}
		\frac{1}{\sqrt{3}} \\ \frac{1}{\sqrt{6}}e^{-i\varphi_k} \\ \frac{1}{\sqrt{6}}e^{-i\varphi_k} \\
		\frac{1}{\sqrt{3}}e^{-i\phi_k}
	\end{pmatrix} \ k=1,2,3.  $$ 
	With $\{ \varphi_k \} $ and $\{ \phi_k \} $ form two independent equilateral triangles.
 
    This result seems reasonable because we get one free parameter solution in the one-qubit problem and the solution with two free parameters in the two copies of the qubit case. The symmetric relations of the free parameter are both equilateral triangles. The more copies state case is discussed in section \ref{result2}.

    \subsection{Nontrivial optimal POVM: two-parameter qubit} \label{nontrivial}
    This section is an extension of section \ref{qubittwoparameters} and describes the two-parameter qubit state at an arbitrary point in the parameter space.
    \begin{center}
    \begin{tabular}{@{}lll}
    \br
     state $\rho_\theta$   & derivatives $ \partial_i \rho$  & $ K^*$  \\
    \mr
     $\frac{1}{2} \left( I+\theta_1 \sigma_1 +\theta_2 \sigma_2 \right)
     $   &   $ \partial_1 \rho =\frac{1}{2}\sigma_1 $, $ \partial_2 \rho =\frac{1}{2} \sigma_2 $ 
     &   3   \\
    \br
    \end{tabular}    
    \end{center}
Here $\theta=(\theta_1,\theta_2)$ is fixed but arbitrary. In other words, the true state is not at the origin in the Bloch vector space but all the feasible $\theta$ with $||\theta|| \leq 1$. It is expected that the optimal POVM will depend on $\theta$. With a simple calculation, it is derived that there is no factor on $\sigma_3$ as measuring the $z$ component does not give any information about the parameters. The optimal POVM with minimal outcome is three. Firstly, we introduce the previously known optimal one with four outcomes.
    
\subsubsection{Optimal measurement with randomized PVM}
~\\
This optimal measurement is given by the structure of randomized PVM \cite{yamagata2011}. Let $\theta= ( r \cos \varphi,r \sin \varphi,0)$, the unit vector among $\theta$ direction would be $\Vec{n}^{(1)}=\frac{\theta}{r}=(\cos\varphi,\sin\varphi,0)$. The perpendicular vector is defined for any unit vector orthogonal to $\theta$. For instance, $\vec{n}^{(2)}=\frac{\theta^\perp}{r}=(-\sin\varphi,\cos\varphi,0)$. Then the pair of projection measurements for each direction is defined as
\begin{align*}
\Pi_{\pm}^{(i)}=\frac{1}{2}\left( I\pm \vec{n}^{(i)} \cdot \vec{\sigma} \right) \,\quad \mathrm{for }\ i=1,2, 
\end{align*}
where $\vec{n}\cdot \Vec{\sigma}=n_1 \sigma_1+n_2 \sigma_2+n_3 \sigma_3 $ and $\vec{\sigma}= (\sigma_1,\sigma_2,\sigma_3) $, the vector of Pauli matrices. As a result, the optimal measurement is defined by the randomized combination,
\begin{align*}
\Pi &= p_1 \Pi^{(1)} \oplus p_2 \Pi^{(2)} \\
& = \left( p_1 \Pi_{+}^{(1)},p_1 \Pi_{-}^{(1)},p_2 \Pi_{+}^{(2)},p_2 \Pi_{-}^{(2)} \right) ,   
\end{align*}
with $p_1=\frac{\sqrt{1-r^2}}{\sqrt{1-r^2}+1 }, p_2=\frac{1}{\sqrt{1-r^2}+1 }.$
This measurement is indeed optimal by straightforwardly substituting it and confirming the equality. By looking at the formula of this measurement, we notice that this is the solution with four outcomes which is not the minimal number. The optimal one with three outcomes (minimal number) is given in the next part.

\subsubsection{Optimal measurement with three outcomes}
\begin{figure}[!htb]
   \begin{minipage}{0.48\textwidth}
     \centering
     \includegraphics[width=\linewidth,page=3]{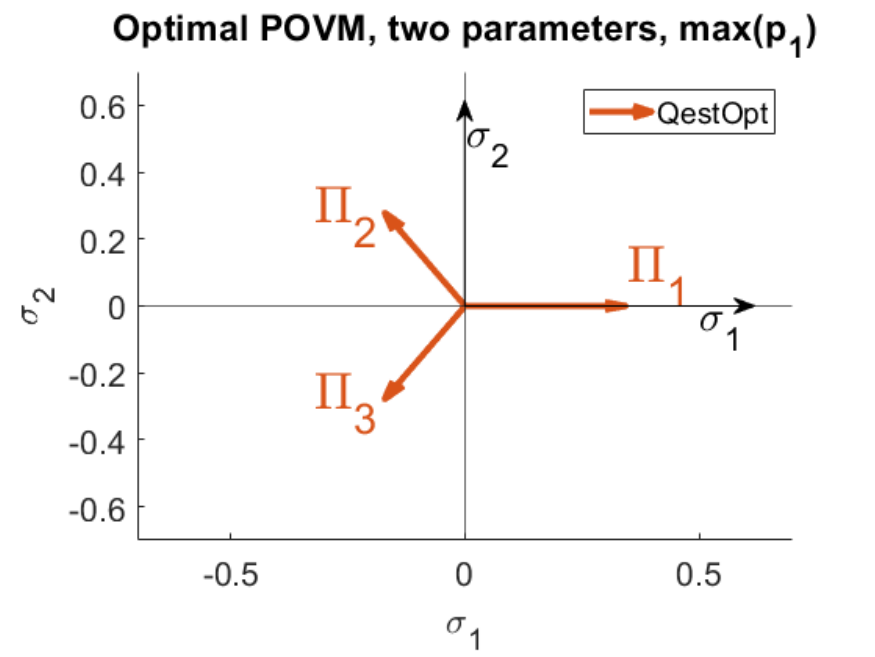}
   \end{minipage}\hfill
   \begin{minipage}{0.48\textwidth}
     \centering
     \includegraphics[width=\linewidth,page=4]{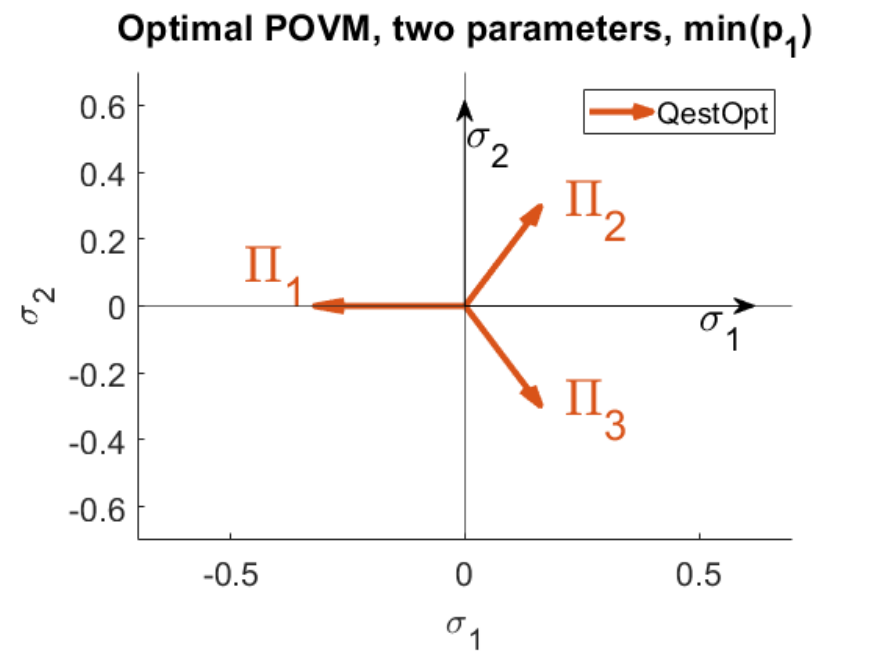}
   \end{minipage}
   \caption{The coefficients of the optimal POVM on the directions of $\sigma_1$ and $\sigma_2$ when the absolute value of parameters, $r=0.1$ forms in three arrows. The probability of the first POVM elements, $p_1$ takes the maximal and minimal values respectively. The model is the two-parameter Bloch model with an arbitrarily fixed state.}\label{Fig:Data1}
\end{figure}

\begin{figure}[!htb]
   \begin{minipage}{0.48\textwidth}
     \centering
     \includegraphics[width=\linewidth,page=5]{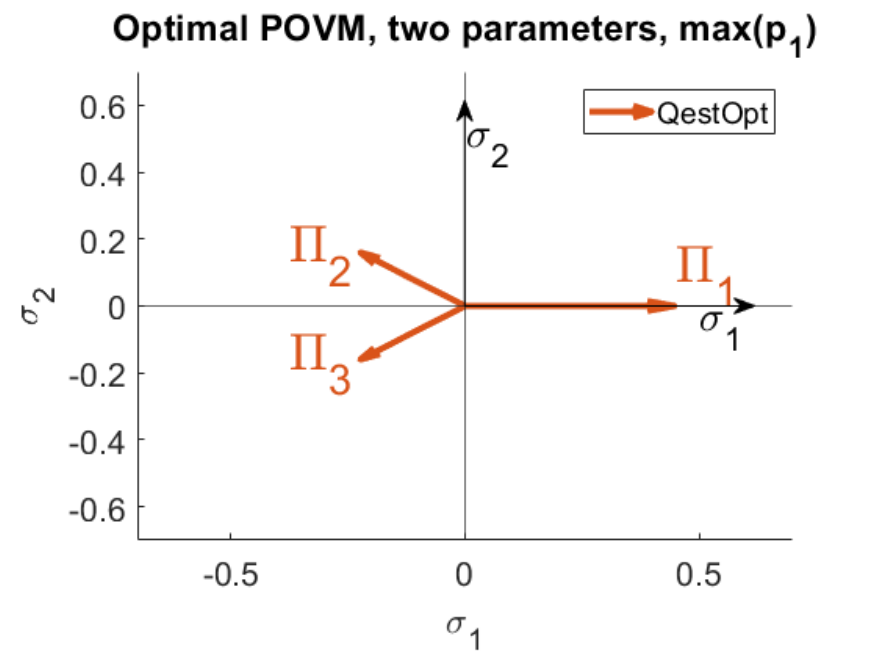}
   \end{minipage}\hfill
   \begin{minipage}{0.48\textwidth}
     \centering
     \includegraphics[width=\linewidth,page=6]{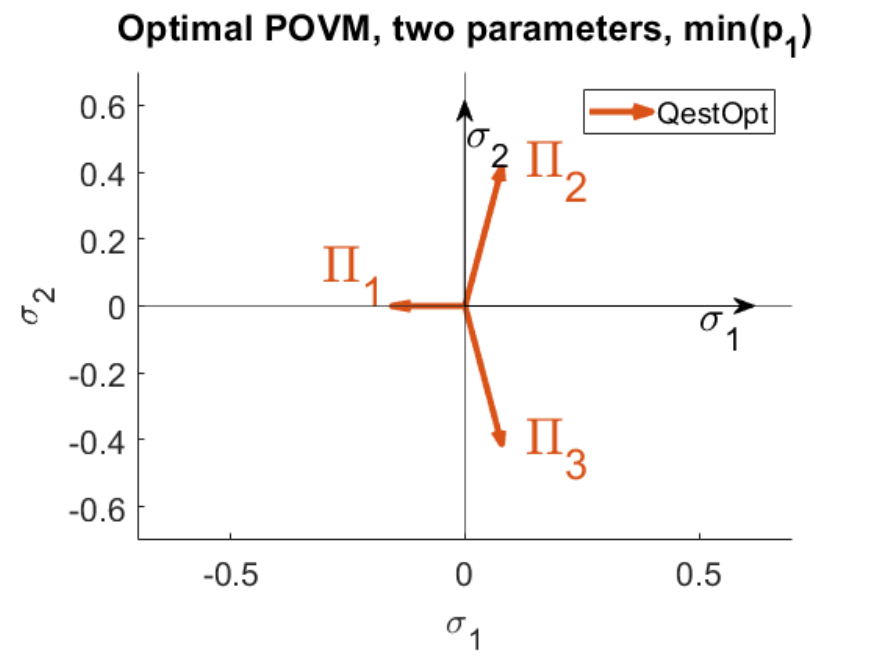}
   \end{minipage}
   \caption{The coefficients of the optimal POVM on the directions of $\sigma_1$ and $\sigma_2$ when the absolute value of parameters, $r=0.9$ forms in three arrows. The probability of the first POVM elements, $p_1$ takes the maximal and minimal values respectively. The model is the two-parameter Bloch model with an arbitrarily fixed state.}\label{Fig:Data2}
\end{figure}

~\\
Since the component on $\sigma_3$ does not give any information about the parameters, we can parameterize the optimal POVM as 
\begin{equation}\label{threeoutcomes}
    \Pi_k=p_k\left(I+\cos\phi_k \sigma_1  +\sin\phi_k \sigma_2  \right) ~,~ k=1,2,3.
\end{equation} If we denote the length of the Bloch vector by $r=||\theta||=\sqrt{\theta_1^2+\theta_2^2}$, the proportionality among the three numbers $p_1:p_2:p_3$ depends on the length $r$.  In a special case, we notice that if $\theta$ is close to the origin, three $p_k$s will be approximate to $\frac{1}{3}$. This means $p_1=p_2=p_3$, and the form reduces to the section \ref{qubittwoparameters}. Generally speaking,
for arbitrary $\theta=(r \cos \varphi, r\sin \varphi)$, we prove that the optimal POVM with three outcomes explicitly forms in (\ref{threeoutcomes}),
where $p_k$ is chosen from one-free-parameter equations,
	\begin{align*}
		\sum_{k=1}^3 p_k &= 1 ,\\
		\sum_{k=1}^3 \frac{1}{1-2p_k} &= 5+ \frac{4}{\sqrt{1-r^2}}.
	\end{align*}			
The $\phi_k$ is determined by
\begin{align*} \phi_k=\arccos \left\{ \frac{1}{2r} \left[ \frac{\sqrt{1-r^2}}{1-2 p_k} - (2+\sqrt{1-r^2}) \right] \right\} .\end{align*}
The $\sin \phi_k$ is determined by (C.3). The details of the proof are written in Appendix C. The solution is not unique since there are only two conditions for $p_1,p_2,p_3$. It is determined if we add another condition or consider the minimal value of any $p_k$. For example, \figtext \ref{Fig:Data1} and \figtext \ref{Fig:Data2} show the optimal POVMs with minimal $p_1$ and maximal $p_1$ for $r=0.1$ and $r=0.9$. The $xy$ axes in \figtext \ref{Fig:Data1} and \ref{Fig:Data2} are coefficients on $\sigma_1$ and $\sigma_2$. For $r \ll 1 $, we have an approximately equilateral one and the optimal POVM degenerates in the form of $\theta$ at the origin case.

\section{Result2: Multiple copies of qubit state} \label{result2}

    Performing collective measurements on i.i.d. qubits, involving the simultaneous estimation of multiple qubits, offers the advantage of reducing the MSE. A specific illustration of this advantage has been presented in section \ref{qubittwoparatwocopy} for the case of two qubits. The subsequent discussion in this section extends the application of this principle to scenarios involving more than two qubits. The algorithm introduced here is tailored to find the optimal POVM for any given number of qubit copies, contingent on the computational capabilities of the used computer. In our simulation, we have achieved optimal results for up to six qubit copies for two-parameter models due to the time limitation.

    The state we are handling is the Bloch vector parametrization of a qubit state: $\rho=\frac{1}{2}\left( I+\theta_1\sigma_1+\theta_2\sigma_2+\theta_3\sigma_3  \right)$, where $\sigma_i$ are the Pauli matrices. We will consider three variants of this model. The partial derivative of state concerning $\theta_i$ is given by $\partial_i \rho=\frac{1}{2} \sigma_i$. The i.i.d copies are written in tensorial form,
    \begin{align*} \rho^{\otimes M}=\underbrace{\rho \otimes \rho \otimes \ldots \otimes \rho \otimes \rho}_M, \end{align*}
    where $M$ is the number of copies. Under these settings, the derivative of the i.i.d. copies is
    \begin{align*}
    \partial_i (\rho^{\otimes M})=\partial_i \rho \otimes \rho \otimes \ldots \otimes \rho+\rho \otimes \partial_i \rho \otimes \ldots \otimes \rho+\ldots + \rho \otimes \rho \otimes \ldots \otimes \partial_i \rho .
    \end{align*}

    \subsection{Multiple copies of two-parameter qubit}
First, we consider the case $\theta_3$ is known, and it is fixed as $\theta_3 =0$. In the following, we find an optimal POVM for $M=2,3,4,5,6$ qubit cases by our algorithm, and then as we described in section \ref{lowerbound}, compared with the Nagaoka bound. Since we are interested in the best accuracy for point estimation, we need to fix a point to be estimated. After changing the point in the Bloch sphere $\{ \theta_1^2+\theta_2^2\leq 1 \}$, we could immediately find that the objective value is rotationally symmetric with $\theta_1, \theta_2  $. In other words, the value only depends on the distance from $(\theta_1,\theta_2) $ to the origin. Due to this, we vary $ \theta_1 $ from 0 to 1 whereas another parameter is set to $\theta_2=0$. The Holevo bound in this case is equal to the SLD bound.
    

\begin{figure}[!htb]
     \centering
     \includegraphics[width=0.48\linewidth,page=7]{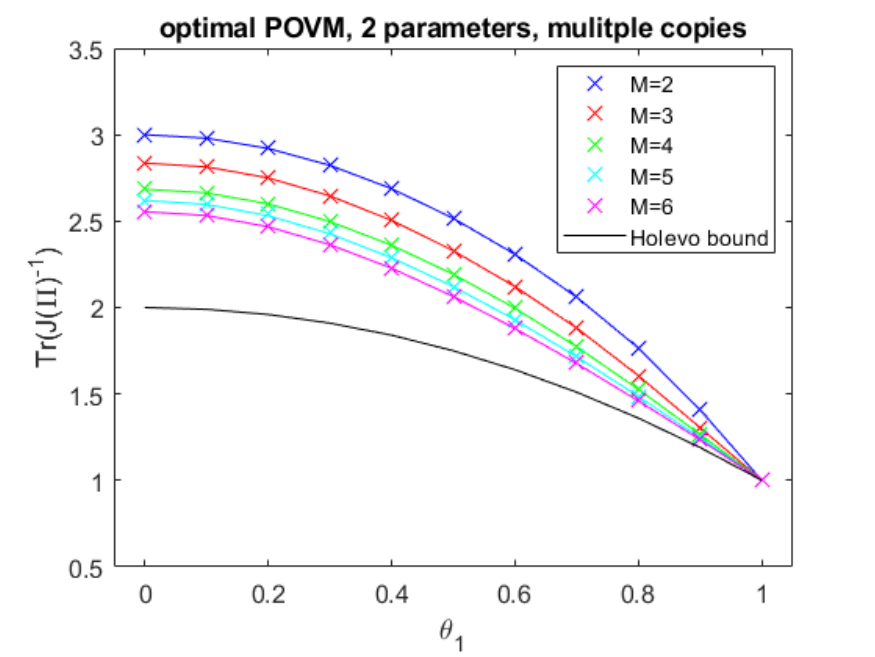}
     \caption{Coefficient on the direction of $\sigma_1$( $\theta_1$) versus $\Tr(J(\Pi^*)^{-1})$(cross point), the NH bound (five curves from above) and the Holevo bound (the lowest curve). $\Pi^*$ is found by sufficient iterations in QestOptPOVM. The NH bound is found by the SDP solver. The Holevo bound is found by direct calculation. The model is the two-parameter for multiple copies as $M=2,3,\ldots,6$.}
   \label{fig:multi2}
\end{figure}

We plot $\theta_1$ versus $\min_{\mathrm{QestOptPOVM}} \Tr(J(\Pi)^{-1})$ in \figtext \ref{fig:multi2}. The cross presents the minimal value of the objective function $\Tr(J^{-1}(\Pi^*))$, whiich is found by QestOptPOVM after sufficiently large number of iterations. The five curves from above show the NH bound from different numbers of copies ($M=2,3,\ldots,6$). As a comparison, the Holevo bound (=the SLD bound) is shown in the black curve. When the line passes through the center of the cross, we can say that the minimal objective value is equal to the NH bound. The NH bound is attained by the POVM found in this algorithm. This means, on the one hand, this Nagaoka bound is attainable, on the other hand, this POVM is the optimal one. 

The steps to get one point in \figtext \ref{fig:multi2}:
Enumerate $\theta_1$ from 0 to 1. For each $\theta_1$, use the algorithm QestOptPOVM to find optimal measurements. The algorithm starts with a randomly chosen POVM. In these $M=2,3,4$, we obtain the data as in \figtext \ref{fig:multi2} by arbitrary initial random POVM. This shows the efficiency of our algorithm. Since the result depends on the random initialization, for a difficult problem, namely $M\geq5$, it cannot avoid local minima. We run it 100 times with different initials and select the minimum one to avoid this problem. The next step is to compute the objective function with that measurement to get one point in the figure.

From \figtext \ref{fig:multi2} we are highly confident that the NH bound is attained by an optimal POVM for all the $\theta_1$ from 0 to 1. Explicit forms for the numerically found optimal POVMs are not shown here, but they are available upon a reasonable request. 
The average gap between the objective value given by these optimal POVMs and the NH bound among all of the parameter values we tried is less than $10^{-6}$. We thus confirm that in these cases, our algorithm is highly accurate and it can find the optimal POVM efficiently. We also report that the minimum number of the optimal POVM elements for $M=2,3,4,5$ are 4,7,10 and 13, respectively, and they are irrespective of the parameter $\theta_1$.

Finally, we observe that the Holevo bound in this model is far below the variance of the optimal POVM even for $M=6$, which is a collective measurement on six copies ($\dim \Hil=2^6=64$). This kind of gap between the Holevo bound and the NH bound was investigated before \cite{conlon2022gap}. However, our algorithm explicitly demonstrates finite gaps based on the exact calculation of the optimal POVMs.

 \subsection{Multiple copies of two-parameter qubit noise model}
Next, consider the case $\theta_3 = 2 \epsilon-1$ is known where $\epsilon$ is fixed. This model is motivated by the dephazing noise along the $z$ axis. We aim at finding optimal POVMs at $(\theta_1,\theta_2) =(0,0)$.
\begin{figure}[!htb]
    \centering
   \includegraphics[width=0.48\linewidth,page=8]{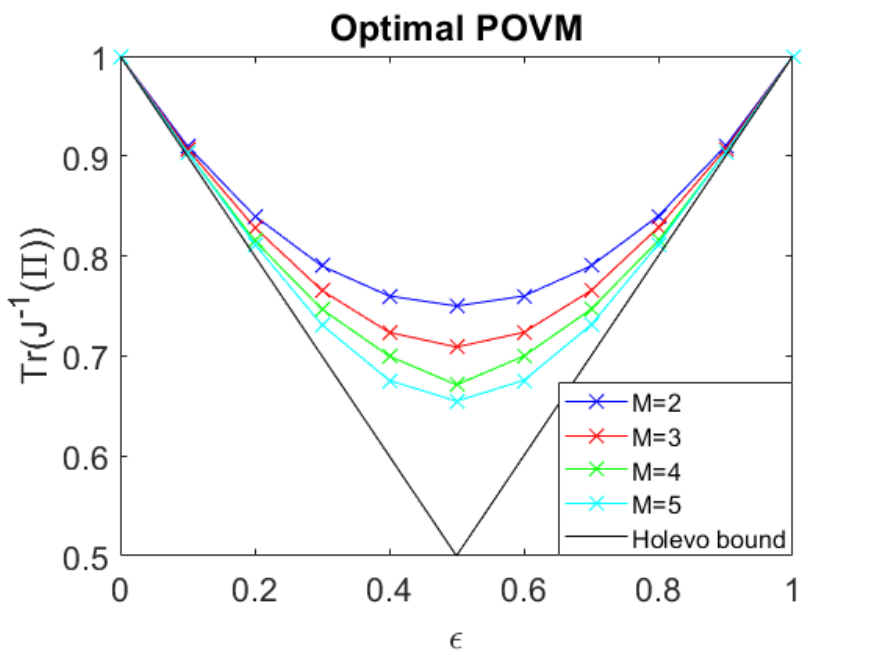}
   \caption{$\epsilon$ versus objective function $\Tr(J(\Pi^*)^{-1})$ (cross) and the Holevo bound (the beneath black line shape in `V') is visualized in the qubit error model for multiple copies. $\Pi^*$ is found by sufficient iterations in QestOptPOVM. The Holevo bound is found by direct calculation. }\label{Fig:eps}
\end{figure} The parametric model is written as
\begin{align*}
\rho(\epsilon)=\begin{pmatrix}
    \epsilon & 0 \\
    0 & 1-\epsilon
\end{pmatrix}.
\end{align*}
In this model, the Holevo bound is equal to the SLD bound.
The optimal value concerning $\epsilon$ for the number of copies from two to five is illustrated in \figtext \ref{Fig:eps}.
The minimal value of the objective function decreases as the number of copies increases because we can extract more information from more copies of a state except the pure state. 
It is clear that the gap between the NH bound and the Holevo bound is inevitably large even in five copies (the $\dim \Hil=2^5=32$) since the NH bound is very close to our optimal value in this model. Within a specific value of copies, the curve is concave up and exhibits $\epsilon=0.5$ symmetry. It reaches its lowest point at $\epsilon=0.5$ which is the most mixed state and it is also the lowest point of the Holevo bound.

 \subsection{Qubit three parameters with multiple copies}\label{qubitthreeparameters}
 \begin{figure}
     \centering
     \includegraphics[width=0.48\linewidth,page=9]{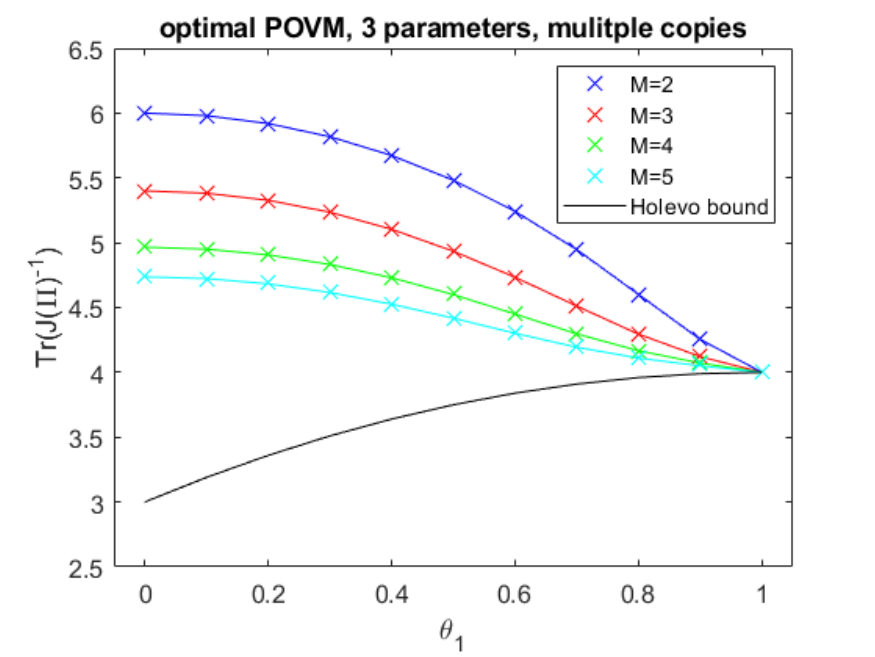}
     \caption{Coefficient on the direction of $\sigma_1$( $\theta_1$) versus $\Tr(J(\Pi^*)^{-1})$(cross point), the NH bound (five curves from above) and the Holevo bound (the lowest curve). $\Pi^*$ is found by sufficient iterations in QestOptPOVM. The NH bound is found by the SDP solver. The Holevo bound is found by direct calculation. The model is the three-parameter for multiple copies as $M=2,3,\ldots,5$.}
     \label{Fig:mult3}
 \end{figure}
This section concerns the three-parameters model for qubit which is the complete parametric space. In this part, finding the numerically optimal POVM is more difficult than the two-parameters case. This is because the dimension of the CFIM increases. However, this difference does not make our algorithm unable to accurately find the optimal value. The accuracy is consistently high up to five copies. By the same reason of two-parameter case, we notice the rotational symmetric with $\theta_1,\theta_2$ and $\theta_3$. We vary $\theta_1$ from 0 to 1 whereas $\theta_2=\theta_3=0$. In \figtext \ref{Fig:mult3}, as before, the lines except the Holevo bound are the NH bound for multiple copies of the state, and the cross is the optimal objective function found by our algorithm. The line passes through the center of the cross with less than $10^{-3}$ errors even in the worst case. This is what we know as the tightness of the NH bound in this model. The efficiency slightly decreases due to the stability of choosing initialization. This is what we can expect as we increase the number of parameters. In particular, for two and three copies, the optimal POVM is still stable to be found by starting with any one random initialization. For the four and five copies, this algorithm needs about one hundred initials to go. At most two hundred initials, QestOptPOVM can find the optimal value for each point in this case. The minimal number of optimal POVMs outcomes for $M=2,3,4,5$ are at most $K^*=5,10,20,40$, respectively.

The Holevo bound in this case is equal to the right logarithmic derivative (RLD) bound \cite{holevobook,YL73}. The gap between the Holevo bound and the NH bound becomes large when $\theta_1\ \simeq 0$ which is more mixed state. By contrast, the gaps shrink when the states are more pure, and they are equal when $\theta_1=1$ which is the pure-state limit \cite{KM}. 


\section{Conclusion} 

We introduce the Quantum Estimation Optimization for POVM (QestOptPOVM) algorithm, a highly efficient and accurate optimization technique designed specifically for optimizing the MSE function in quantum multiparameter estimation. Through the utilization of QestOptPOVM, we have achieved numerically optimal POVM results in the context of qubit systems and multiple copies, with high precision. The space of POVMs we explore encompasses all those not previously considered in other studies. The analytical form derived from our numerical results has been proven to be optimal under sufficient conditions, with one analytical form previously unknown prior to this research.
 
Built upon the steepest gradient descent approach, we have tailored our algorithm to account for the fundamental structure of POVMs space and the nonlinear objective function, which is the trace of the inverse CFIM. Our algorithm ensures monotonicity in each iteration, facilitating a systematic approach towards identifying the optimal POVM. Our algorithm efficiently identifies the optimal POVM in a two-parameter qubit model for up to six i.i.d. states. Despite operating in a $64$-dimensional Hilbert space, the algorithm demonstrates high accuracy, achieving precision up to $10^{-6}$. Even in more challenging scenarios, such as three parameters with five copies, the accuracy remains relatively high at $10^{-3}$ in the worst case. This indicates that QestOptPOVM is an efficient and accurate algorithm in POVM optimization, primarily due to its focused approach on optimizing the MSE over all possible POVMs. Practically, this algorithm provides a tool for evaluating the efficacy of a given POVM. This work not only presents a powerful numerical tool but also paves the way for analytical solutions in previously elusive cases. 

Our study underscores the efficacy of the steepest gradient descent method in POVM optimization while suggesting avenues for future research, such as exploring second-order gradient methods and employing representation theory for higher-dimensional Hilbert spaces. Further stability testing is warranted, and the estimator proposed by our optimal POVM can be compared with standard tomography methods. Additionally, while the analytical form of the optimal POVM in complete Bloch qubit and two-parameter, two-copy qubit scenarios appears solvable, it necessitates further computation.
\section*{Data availability}
The data that support the findings of this study are available from the corresponding author upon reasonable request.
\section *{Acknowledgment}
	The work is partly supported by JSPS KAKENHI Grant Numbers JP21K04919, JP21K11749. JZ is also supported by
	the research assistant scholarship at the University of Electro-Communications.

\onecolumn
\appendix
\section*{Appendix A: the derivative of the main function}
	\renewcommand{\theequation}{A.\arabic{equation}}

    The algorithm is based on gradient descent. Our aim is to find the $\min \Tr (J(\Pi)^{-1})$ where $\Pi=\{\Pi_1,\Pi_2 \ldots \Pi_K \}$, $\Pi_k \geq 0$ and $\sum_k \Pi_k=I$ with a given K. Using the Kraus operator $A$ to replace POVM is a method to keep the positive condition as $A_k^\dag A_k=\Pi_k$. The completeness relation can be formulated in Lagrange multiplier $\Tr (\Lambda ( \sum_k \Pi_k-I ) )$. The equation we need to minimize is 
    \begin{equation}
    f(A,\Lambda)=  \Tr (J(A)^{-1})+  \Tr (\Lambda ( \sum_{k=1}^K A^\dag_k  A_k -I ) )    . 
    \end{equation}
    As utilizing the gradient descent, we set an annotation in the upper right corner with parentheses $A^{(m)} $ to represent the number of iterations. Then we are capable of denoting the update as 
    \begin{equation}
    A_{k}^{(n+1)}=A_{k}^{(m)}+\alpha H_{k}^{(m)} ,
    \end{equation}
    where $\alpha$ is the step size and $H_{k}^{(m)}$ is the update direction. Here we can obtain the derivative of $A_{k}^{(m)}$ by the natural definition of derivative as $\alpha \rightarrow 0$.
    \[
    \left( A_{k}^{(m)}\right)' = H_{k}^{(m)} .
    \]
    Here $'$ means a general derivative. Because $ [\Tr (J(A)^{-1})]'=\Tr ( \left(J(A)^{-1} \right)')=\Tr (-J(A)^{-1} [J(A)]' J(A)^{-1}) $, the item of $[J(A)]'$ is checked firstly. To simplify the equations, several denotations are listed.
    \begin{align*}
        p_k&=\Tr( A_k \rho A_k^\dag), \\
        d_{i,k}&= \Tr( A_k \partial_i \rho A_k^\dag), \\
        D_{i,k}&= \sum_{j=1}^n (J^{-1})_{j,i} d_{j,k}, \\
        \rho^i&= \sum_{j=1}^n (J^{-1})_{j,i} \partial_j \rho , \\
        l_k^i&=\frac{D_k^i}{p_k} .
    \end{align*}
    The $i,j$-th component of the derivative of $[J(A)]'$ is
    \begin{align*}
            ([J(A)]')_{i,j}&=\left[ \sum_{k=1}^K \frac{\partial_i p (x|A) \partial_j p (x|A) }{p (x|A) } \right]'\\
        &= \sum_{k=1}^K \left[ \frac{d_{i,k} d_{j,k} }{p_k } \right]'\\
        &= \sum_{k=1}^K \left[ \frac{d_{i,k}' d_{j,k} }{p_k }
        +\frac{d_{i,k} d_{j,k}' }{p_k }
        -\frac{d_{i,k} d_{j,k} }{p_k^2 } p_k'\right] .
    \end{align*}
    Compute $n+1$-th iteration $p_k' $ and $d'_{i,k}$ separately,
    \begin{align*}
    p_k'&=[\Tr(A_k \rho A_k^\dag)]'=\Tr[(A_k^{(m)} \rho A_k^{(m)\dag} )']=\Tr[H_k^{(m)} \rho A_k^{(m)\dag}+A_k^{ (m) } \rho H_k^{(m)\dag} ] \\
    &=\Tr[H_k^{(m)} \rho A_k^{(m)\dag}+A_k^{ (m) } \rho H_k^{(m)\dag} ],
    \end{align*}
    when $\alpha \rightarrow 0$. Similarly we get the form of $d'_{i,k}$
    \[
    d_{i,k}'=\Tr[H_k^{(m)} \partial_i \rho A_k^{(m)\dag}+A_k^{ (m) } \partial_i \rho H_k^{(m)\dag} ] .
    \]
    Substitute these to calculate the $i,j$ item of derivative of Fisher information obtaining
    \begin{align*}
        ([J(A)]')_{i,j}&=\sum_{k=1}^K \Tr \left[ H_k^{(m)} X_{k,i,j}^{(m)\dag}A_k^{(m)\dag}
        +A_k^{(m)} X_{k,i,j}^{(m)}H_k^{(m)\dag}\right],
    \end{align*}
    where $X_{k,i,j}^{(m)}= \frac{\partial_i \rho d_{j,k}}{p_k}+\frac{\partial_j \rho d_{i,k}}{p_k}-\frac{\rho d_{i,k} d_{j,k}}{p_k^2} $. The derivative of $\Tr(J(A)^{-1})$ will be
    \begin{align*}
        [\Tr (J(A^{(m)})^{-1})]' &= - \Tr (J(A^{(m)})^{-1} [J(A^{(m)})]' J(A^{(m)})^{-1})\\
        &=- \sum_{\ell=1}^n\sum_{i=1}^n\sum_{j=1}^n J^{\ell,i} [J(A^{(m)})]'_{i,j} J^{j,\ell}\\
        &=- \sum_{\ell,i,j=1}^n  J^{\ell,i} \sum_{k=1}^K \Tr \left[ H_k^{(m)} X_{k,i,j}^{(m)\dag}A_k^{(m)\dag}
        +A_k^{(m)} X_{k,i,j}^{(m)}H_k^{(m)\dag}\right] J^{j,\ell}\\
        &=- \sum_{k=1}^K \sum_{\ell,i,j=1}^n   \Tr \left[ H_k^{(m)} J^{\ell,i}  X_{k,i,j}^{(m)\dag} J^{j,\ell} A_k^{(m)\dag}
        +A_k^{(m)} J^{\ell,i} X_{k,i,j}^{(m)} J^{j,\ell}  H_k^{(m)\dag}\right]\\
        &=- \sum_{k=1}^K   \Tr \left[ H_k^{(m)}   X_{k,}^{(m)\dag} A_k^{(m)\dag}
        +A_k^{(m)}  X_{k}^{(m)}  H_k^{(m)\dag}\right], 
    \end{align*}
    where $ J^{\ell,i}=[J(A)^{-1}]_{\ell,i} $ and $X_k^{(m)}=\sum_{\ell,i,j=1}^n J^{\ell,i} X_{k,i,j}^{(m)} J^{j,\ell}$. The remaining item is the Lagrange multiplier. The derivative of it would be
    \begin{align*}
         [\Tr (\Lambda^{(m)} ( \sum_{k=1}^K A^{(m)\dag}_k  A^{(m)}_k -I ) ) ]'
         &=  \sum_{k=1}^K \Tr  \left[ \Lambda^{(m)} (A^{(m)\dag}_k  A^{(m)}_k -I )' \right] \\
         &= \sum_{k=1}^K \Tr  \left[ \Lambda^{(m)} (H^{(m)\dag}_k  A^{(m)}_k + 
         A^{(m)\dag}_k H^{(m)}_k   ) \right]\\
         &= \sum_{k=1}^K \Tr  \left[ H^{(m)}_k \Lambda^{(m)}  A^{(m)\dag}_k  + A^{(m)}_k  \Lambda^{(m)} H^{(m)\dag}_k     \right].
    \end{align*}
Combine two items of the derivative of $f$ to get the total form of $f'$ in
    \begin{align*}
        f'(A^{(m)},\Lambda^{(m)}) =& - \sum_{k=1}^K   \Tr \left[ H_k^{(m)}   X_{k,}^{(m)\dag} A_k^{(m)\dag}        +A_k^{(m)}  X_{k}^{(m)}  H_k^{(m)\dag}\right] 
        \\ &  +\sum_{k=1}^K \Tr  \left[ H^{(m)}_k \Lambda^{(m)}  A^{(m)\dag}_k  + A^{(m)}_k  \Lambda^{(m)} H^{(m)\dag}_k     \right]\\
        =& - \sum_{k=1}^K \Tr  \left[ H^{(m)}_k \left(  X^{(m)\dag}_k -\Lambda^{(m)} \right) A^{(m)\dag}_k  + A^{(m)}_k \left( X^{(m)}_k -\Lambda^{(m)} \right)  H^{(m)\dag}_k     \right].
    \end{align*}
    This form inspire to let the update direction matrix be $H^{(m)}_k=A^{(m)}_k \left( X^{(m)}_k -\Lambda^{(m)} \right)$.
    The derivative of $f$ becomes
    \begin{align*}
    f'(A^{(m)},\Lambda^{(m)}) &=- 2 \sum_{k=1}^K \Tr\left[ H^{(m)}_k H^{(m)\dag}_k  \right]\\
    &= -2 \sum_{k=1}^K \langle H^{(m)}_k,H^{(m)}_k \rangle_{HS} \leq 0.
    \end{align*}
    This means that as $\alpha$ close to zero, this choice of $H^{(m)}_k$ ensures non-positivity for the derivative.
    The next step is to solve the Lagrange multiplier. The condition of taking the derivative of $\Lambda$ would be the completeness relationship, 
    $\sum_{k=1}^K A_k^\dag A_k = I $. Furthermore, we require the left-hand side to still be equal to the identity after the update. This means the general derivative of the left-hand side is zero.    
    \begin{align*}
        \left( \sum_{k=1}^K A_k^\dag A_k  \right)'&=0,\\
        \sum_{k=1}^K \left( H_k^\dag A_k + A_k^\dag H_k \right) &=0,\\
        \sum_{k=1}^K \left( \left( X^{(m)\dag}_k -\Lambda^{(m)} \right) A_k^{(m)\dag} A^{(m)}_k + A_k^{(m)\dag} A^{(m)}_k \left( X^{(m)}_k -\Lambda^{(m)} \right) \right) &=0,\\
        \sum_{k=1}^K \left(  X^{(m)\dag}_k  A_k^{(m)\dag} A^{(m)}_k -\Lambda^{(m)}  A_k^{(m)\dag} A^{(m)}_k + A_k^{(m)\dag} A^{(m)}_k  X^{(m)}_k -A_k^{(m)\dag} A^{(m)}_k  \Lambda^{(m)}  \right) &=0,\\
        \sum_{k=1}^K \left(  X^{(m)\dag}_k  A_k^{(m)\dag} A^{(m)}_k + A_k^{(m)\dag} A^{(m)}_k  X^{(m)}_k  \right) &=2 \Lambda^{(m)}. \\
    \end{align*}
    This implies that 
    \[
     \Lambda^{(m)} = \frac{1}{2} \sum_{k=1}^K \left(  X^{(m)\dag}_k  A_k^{(m)\dag} A^{(m)}_k + A_k^{(m)\dag} A^{(m)}_k  X^{(m)}_k  \right).
    \]
    In summary, the iteration item would be:
    \begin{equation} \label{Akn}
    \begin{split}
        A^{(m+1)}_k &= A^{(m)}_k+\alpha H^{(m)}_k\\
        &=  A^{(m)}_k+\alpha  A^{(m)}_k \left( X^{(m)}_k - \Lambda^{(m)} \right)\\
        &=  A^{(m)}_k\left( I+ \alpha (  X^{(m)}_k - \Lambda^{(m)})\right),
    \end{split}
    \end{equation}
    where  
    \begin{equation} \label{Xkn}
        \begin{split}
        X_k^{(m)}&=\sum_{\ell,i,j=1}^n J^{\ell,i} X_{k,i,j}^{(m)} J^{j,\ell}\\
        &=\sum_{\ell,i,j=1}^n J^{\ell,i} \left(
        \frac{\partial_i \rho d_{j,k}}{p_k}+\frac{\partial_j \rho d_{i,k}}{p_k}-\frac{\rho d_{i,k} d_{j,k}}{p_k^2} \right) J^{j,\ell} \\
        &=\sum_{\ell=1}^n  \left(
        \frac{ \rho^\ell D_k^\ell }{p_k}+\frac{ \rho^\ell D_k^\ell}{p_k}-\frac{\rho D^\ell_k D^\ell_{k}}{p_k^2} \right)  \\
        &=\sum_{\ell=1}^n  \left(
        2 \rho^\ell l_k^\ell - \rho (l^\ell_k)^2  \right),  \\
        \end{split}
    \end{equation}
    and 
    \begin{equation}
    \Lambda^{(m)} = \frac{1}{2} \sum_{k=1}^K \left(  X^{(m)\dag}_k  A_k^{(m)\dag} A^{(m)}_k + A_k^{(m)\dag} A^{(m)}_k  X^{(m)}_k  \right) .
    \end{equation}

\section*{Appendix B: Qubit three parameter optimality proof}

Proof of necessary condition:
	
    The optimal POVM can be specified by any unit tetrahedron. Since the tetrahedron in $xyz$-axis is a three free variables space, we can use Euler angles to represent the form of POVM.

    The optimal POVMs are: $\Pi_k=\frac{1}{4}\left( I+RV_k\cdot \vec{\sigma} \right)$, $k=1,2,3,4$, where
    $V_1=\frac{\sqrt{3}}{3}(1,1,1)^T,V_2=\frac{\sqrt{3}}{3}(1,-1,-1)^T,V_3=\frac{\sqrt{3}}{3}(-1,1,-1)^T,V_4=\frac{\sqrt{3}}{3}(-1,-1,1)^T$ are unit tetrahedron, and $RV_1,\ RV_2,\ RV_3,\ RV_4 $ are unit tetrahedron with rotation matrix $R$ that
	\[\begin{split}
		R(\alpha,\beta,\gamma)=R_1(\alpha)R_2(\beta)R_3(\gamma) &=
		\begin{pmatrix} \cos\alpha & -\sin\alpha & 0 \\
			\sin\alpha & \cos\alpha & 0 \\
			0 & 0 & 1 \end{pmatrix}
		\begin{pmatrix} \cos\beta & 0 & \sin\beta \\
			0 & 1 & 0 \\
			-\sin\beta & 0 & \cos\beta \end{pmatrix}
		\begin{pmatrix} 1 & 0 & 0 \\
			0 & \cos\gamma & -\sin\gamma \\
			0 & \sin\gamma & \cos\gamma \end{pmatrix}\\
	\end{split}\]
     \[=\begin{pmatrix} \cos\beta\cos\gamma & \sin\alpha\sin\beta\cos\gamma-\cos\alpha\sin\gamma & \cos\alpha\sin\beta\cos\gamma+\sin\alpha\sin\gamma \\
			\cos\beta\sin\gamma & \sin\alpha\sin\beta\sin\gamma+\cos\alpha\cos\gamma & \cos\alpha\sin\beta\sin\gamma-\sin\alpha\cos\gamma \\
			-\sin\beta & \sin\alpha\cos\beta & \cos\alpha\cos\beta \end{pmatrix}.\]
   Here $\alpha,\beta,\gamma$ are three free variables. Any values of them can derive an optimal POVM. We notice
	\begin{align*}
		\Tr(\partial_i \rho\Pi_k)&=\frac{1}{4} \Tr\left( \sigma_i (I+RV_k\cdot \vec{\sigma}) \right)\\
		&=\frac{1}{4}\cdot 2(RV_k)_i\\
		&=\frac{1}{2}(RV_k)_i .
	\end{align*}
	Then we have $RV_k(RV_k)^T=RV_k V_k^T R^T=\left[(RV_k)_i(RV_k)_j\right]_{i,j}$. 
	With this, we can show
	\begin{align*}
		J(\Pi) &=\left[\sum_k\frac{d^i_k d^j_k}{p_k}\right]_{i,j}\\
		&=\left[\sum_k Tr(\partial_i \rho \Pi_k)Tr(\partial_j \rho \Pi_k) Tr(\rho \Pi_k)^{-1}\right]_{i,j}\\
		&=4\left[\sum_k \frac{1}{2}(RV_k)_i \frac{1}{2}(RV_k)_j \right]_{i,j}\\
		&=\sum_k RV_k V_k^T R^T\\
		&= R \left( \sum_k V_k V_k^T \right) R^T\\
		&= \frac{1}{3} R\left( \begin{pmatrix}1 & 1 & 1 \\
			1 & 1 & 1 \\
			1 & 1 & 1 \end{pmatrix}+\begin{pmatrix}1 & -1 & -1 \\
			-1 & 1 & 1 \\
			-1 & 1 & 1 \end{pmatrix}+\begin{pmatrix}1 & -1 & 1 \\
			-1 & 1 & -1 \\
			1 & -1 & 1 \end{pmatrix}+\begin{pmatrix}1 & 1 & -1 \\
			1 & 1 & -1 \\
			-1 & -1 & 1 \end{pmatrix} \right) R^T\\
		&=\frac{4}{3} R I R^T\\
		&=\frac{4}{3}I .
	\end{align*}
	Therefore, $J^{-1}(\Pi)=\frac{3}{4}I$, and $\Tr \left(J^{-1}(\Pi)\right)=\frac{9}{4}$.

	\section*{Appendix C: Proof of optimal POVM with three outcomes for two parameters state}
Firstly we explain that if we get an optimal measurement with three outcomes on one axis as $\theta = (r,0)$, by performing rotation transformation we obtain the optimal one with $\theta=(\theta_1,\theta_2)= ( r \cos \varphi,r \sin \varphi)$.
We parameterize the POVM as $\Pi_k=p_k\left(I+\cos\phi_k \sigma_1  +\sin\phi_k \sigma_2  \right)$,
    $\Pi:\mathrm{POVM} \Longleftrightarrow
    \sum_k p_k=1 \ \mbox{and}\ \sum_k e^{i\phi_k} p_k=0  $
	then the optimal condition (2) is written as follows. Let $C(r)=\frac{1}{1+\sqrt{1-r^2}} \frac{1}{r^2} $, then
     \begin{align*}
    &\sum_{k=1}^3 \frac{p_k}{1+\theta_1 \cos\phi_k+\theta_2 \sin\phi_k } \begin{pmatrix}
	\cos \phi_k \\ \sin \phi_k  \end{pmatrix} \begin{pmatrix}
	\cos \phi_k  & \sin \phi_k  \end{pmatrix} \\
    =& C(r) \begin{pmatrix}
    \theta_1^2 /\sqrt{1-r^2} +\theta_2^2	 & (1/\sqrt{1-r^2}-1)\theta_1\theta_2  \\
	(1/\sqrt{1-r^2}-1)\theta_1\theta_2 & \theta_2^2 /\sqrt{1-r^2} +\theta_1^2  \end{pmatrix}.
     \end{align*}
    Define $D(r)=\frac{1-\sqrt{1-r^2}}{1+\sqrt{1-r^2}} $, we have two equations:
    \begin{equation}
	\left\{
	\begin{aligned}
		\sum_k \frac{p_k}{1+ r \cos (\varphi -\phi_k)} &= \frac{1}{\sqrt{1-r^2}}, \\
		\sum_k \frac{p_k}{1+ r \cos (\varphi -\phi_k) } e^{2i \phi_k} &= \frac{1}{\sqrt{1-r^2} } D(r) e^{ 2i \varphi}. \\
	\end{aligned}	\tag{$\appn.1$}\label{eq:C1}			
	\right.
\end{equation}
Summarizing the conditions of POVM and $(\appn.1)$ and we get the following equations:
\begin{equation}
	\left\{
	\begin{aligned}
		\sum_{k=1}^3 p_k &= 1, \\
        \sum_{k=1}^3 e^{i( \phi_k-\varphi) } p_k &= 0,\\
        \sum_{k=1}^3 \frac{p_k}{1+ r \cos (\phi_k -\varphi )} &= \frac{1}{\sqrt{1-r^2}}, \\
		\sum_{k=1}^3 \frac{p_k}{1+ r \cos (\phi_k -\varphi ) } e^{2 i(\phi_k -\varphi )} &=\frac{1}{\sqrt{1-r^2} } D(r).
	\end{aligned}	\tag{$\appn.2$ }			
	\right.
\end{equation}
These equations $(\appn.2)$ are satisfied if and only if $p_k,\phi_k$ is the optimal choice for measurement construction. From $(\appn.2)$, it is clear that the angle of $\theta$, namely $\varphi$, contributes in the equations by $(\phi_k - \varphi)$. This means that for any $\theta=(r\cos\varphi, r\sin \varphi)$, it is equivalent to consider $\theta' =(r,0)$ (means $\varphi=0$). After we obtain the optimal $p_k', \phi_k'$ which satisfies $(\appn.2)$ for $\phi'$, we reconstruct $p_k,\phi_k$ by letting $p_k=p_k', \phi_k=\phi_k'+\varphi$. These satisfy $(\appn.2) $ for $\phi$ in straightforward. By this logic, we let $\varphi=0$ without loss of generality. Then the equations become:
\begin{equation}
	\left\{
	\begin{aligned}
		\sum_{k=1}^3 p_k &= 1, \\
        \sum_{k=1}^3 e^{i  \phi_k } p_k &= 0,\\
        \sum_{k=1}^3 \frac{p_k}{1+ r \cos \phi_k } &= \frac{1}{\sqrt{1-r^2}}, \\
		\sum_{k=1}^3 \frac{p_k}{1+ r \cos \phi_k  } e^{2 i \phi_k } &=\frac{1}{\sqrt{1-r^2} } D(r).
	\end{aligned}	\tag{$\appn.3$ }			
	\right.
\end{equation}
To simply the conditions we parameterize $q_k=\frac{p_k \cdot \sqrt{1-r^2}}{1+r\cos\phi_k}$. The good aspect of using this parameterization is that the first equation and third equation in $(\appn.3)$ merge as one. Transformation $(p_k,\phi_k) \Leftrightarrow (q_k,\phi_k)$ is one-to-one by 
\[p_k=\frac{ q_k ( 1+r\cos\phi_k) }{\sqrt{1-r^2}  }, \tag{$\appn.4$} \]
then the equations become (denote $\beta=- \frac{r}{1+\sqrt{1-r^2}})$:
\begin{equation}
	\left\{
	\begin{aligned}
		\sum_{k=1}^3 q_k &= 1, \\
        \sum_{k=1}^3 q_k e^{i  \phi_k }& = - \frac{r}{1+\sqrt{1-r^2}} =\beta(r) , \\
		\sum_{k=1}^3 q_k e^{2 i \phi_k } &=\frac{1-\sqrt{1-r^2}}{ 1+\sqrt{1-r^2} } =\beta^2 (r).
	\end{aligned}	\tag{$\appn.5$ }			
	\right.
\end{equation}
$\beta$ is a function of $r$. Using a substitution $z_k=e^{i\phi_k} -\beta$ the equations is simplified as:
\begin{equation}
	\left\{
	\begin{aligned}
		\sum_{k=1}^3 q_k &= 1, \\
        \sum_{k=1}^3 q_k z_k& = 0, \\
		\sum_{k=1}^3 q_k z_k^2 &= 0, \\
        z_k &=e^{i\phi_k} -\beta .
	\end{aligned}	\tag{$\appn.6$ }			
	\right.
\end{equation}
We noticed that the first three equations have high symmetric relation and derived from these equations we obtain the following equation( $(\appn.6.3)-(\appn. 6.2)^2)$:
\[  (z_1-z_2)^2= - \frac{q_3}{q_1 q_2} z_3^2, \]
and this implies 
\[
z_1-z_2= \pm \sqrt{ \frac{q_3}{q_1 q_2}} i z_3.
\]
Substitute with $\appn. 6.2 $ leads to 
\[
z_1 = \frac{1}{ q_1+q_2} \left( -q_3 \pm i q_2 \frac{q_3}{q_1 q_2}  \right) z_3 , \] and similar for $z_2, z_3$. 
Then it yields the following important relation.
\[
\frac{q_1}{1-q_1}|z_1|^2=\frac{q_2}{1-q_2}|z_2|^2=\frac{q_3}{1-q_3}|z_3|^2 .
\]
This gives us the hint to define $t^2=\frac{q_k}{1-q_k}|z_k|^2$, $t\in \R$. By $ z_k =e^{i\phi_k} -\beta $ which is $|z_k|^2=1+\beta^2-2\beta \cos\phi_k$:
\begin{align*}
    \frac{1-q_k}{q_k} t^2 =1+\beta^2-2\beta \cos\phi_k.  \tag{$\appn.7$}
\end{align*}
To solve $t$, multiply $\sum q_k $ on both side implies
\begin{align*}
    t^2=\frac{1-\beta^2}{2}=\frac{\sqrt{1-r^2}}{1+\sqrt{1-r^2}}.
    \tag{$\appn.8$}
\end{align*}
Substitute $(\appn.8)$ in $(\appn.7)$ derives the relation between $\phi_k $ and $r$:
\begin{align*}
    \cos\phi_k &= \frac{1}{2r} \left[ \frac{\sqrt{1-r^2}}{ q_k} - (2+\sqrt{1-r^2}) \right]. \tag{$\appn.9$}
\end{align*}
Combining $(\appn.4)$ and $(\appn.9)$ induces the relation between $p_k$ and $q_k$:
\begin{align*}
    q_k=1-2p_k. \tag{$\appn. 10$}
\end{align*}
Substitute all in real part of $(\appn. 5.3) $ results in:
\begin{align*}
     \sum q_k \cos 2\phi_k  = \beta^2 
   \Leftrightarrow  \sum \frac{1}{q_k}=\frac{9-\beta^2}{1-\beta^2}.
\end{align*}
Then there are the only remaining two constraints for the equations:
\begin{equation*}
	\left\{
	\begin{aligned}
		\sum_{k=1}^3 q_k &= 1 ,\\
		\sum_{k=1}^3 \frac{1}{q_k} &= \frac{9-\beta^2}{1-\beta^2}.
	\end{aligned}		\tag{$\appn.11$}		
	\right.
\end{equation*}
This is one free parameter constraint because there are three parameters with two equations. Up to now, we have completed the construction of optimal POVMs with three outcomes. For any given $r$, after choosing any $q_1,q_2,q_3$ satisfying $(\appn.11)$, we determine $\cos \phi_k$ by $(\appn.9)$ and $p_k$ by $(\appn.10)$. The last step is to determine the sign of $\sin \phi_k$. Since we have $\sum_{k=1}^3 p_k \sin \phi_k=0$, compare $p_1|\sin\phi_1|$, $p_2|\sin\phi_2|$ and $p_3|\sin\phi_3|$, let the greatest one as positive or negative. The remaining two are negative or positive because you may notice that if $\phi_k$ is the solution of $(\appn.3)$ then $-\phi_k$ is another solution. With the angular transformation, we obtain the $\phi_k+\varphi$.

Finally, we construct the optimal POVM with $\Pi_k=p_k\left(I+\cos(\phi_k+\varphi) \sigma_1  +\sin(\phi_k+\varphi) \sigma_2  \right)$. This optimal measurement with three outcomes found by this specific construction has one free parameter in $(\appn.11)$.

\bibliographystyle{quantum}
\bibliography{myref}

\begin{thebibliography}{10}

\bibitem{kolobov2007quantum}
Mikhail~I Kolobov.
\newblock ``Quantum imaging''.
\newblock \href{https://dx.doi.org/https://doi.org/10.1007/0-387-33988-4}{Springer Science \& Business Media}. ~(2007).

\bibitem{giovannetti2011}
Vittorio Giovannetti, Seth Lloyd, and Lorenzo Maccone.
\newblock ``Advances in quantum metrology''.
\newblock \href{https://dx.doi.org/https://doi.org/10.1038/nphoton.2011.35}{Nature Photonics {\bf 5}, 222}~(2011).

\bibitem{dowling2015quantum}
Jonathan~P Dowling and Kaushik~P Seshadreesan.
\newblock ``Quantum optical technologies for metrology, sensing, and imaging''.
\newblock \href{https://dx.doi.org/10.1109/JLT.2014.2386795}{Journal of Lightwave Technology {\bf 33}, 2359--2370}~(2015).

\bibitem{genovese2016real}
Marco Genovese.
\newblock ``Real applications of quantum imaging''.
\newblock \href{https://dx.doi.org/10.1088/2040-8978/18/7/073002}{Journal of Optics {\bf 18}, 073002}~(2016).

\bibitem{degen17}
Christian~L Degen, F~Reinhard, and P~Cappellaro.
\newblock ``Quantum sensing''.
\newblock \href{https://dx.doi.org/https://doi.org/10.1103/RevModPhys.89.035002}{Reviews of Modern Physics {\bf 89}, 035002}~(2017).

\bibitem{pirandola2018advances}
Stefano Pirandola, B~Roy Bardhan, Tobias Gehring, Christian Weedbrook, and Seth Lloyd.
\newblock ``Advances in photonic quantum sensing''.
\newblock \href{https://dx.doi.org/https://doi.org/10.1038/s41566-018-0301-6}{Nature Photonics {\bf 12}, 724--733}~(2018).

\bibitem{Albarelli2020}
Francesco Albarelli, Marco Barbieri, Marco~G. Genoni, and Ilaria Gianani.
\newblock ``A perspective on multiparameter quantum metrology: From theoretical tools to applications in quantum imaging''.
\newblock \href{https://dx.doi.org/https://doi.org/10.1016/j.physleta.2020.126311}{Physics Letters A {\bf 384}, 126311}~(2020).

\bibitem{rodriguez2021efficient}
Marco~A Rodr{\'\i}guez-Garc{\'\i}a, Isaac~P{\'e}rez Castillo, and P~Barberis-Blostein.
\newblock ``Efficient qubit phase estimation using adaptive measurements''.
\newblock \href{https://dx.doi.org/https://doi.org/10.22331/q-2021-06-04-467}{Quantum {\bf 5}, 467}~(2021).

\bibitem{aasi2013enhanced}
Junaid Aasi, Joan Abadie, BP~Abbott, Richard Abbott, TD~Abbott, MR~Abernathy, Carl Adams, Thomas Adams, Paolo Addesso, RX~Adhikari, et~al.
\newblock ``Enhanced sensitivity of the ligo gravitational wave detector by using squeezed states of light''.
\newblock \href{https://dx.doi.org/https://doi.org/10.1038/nphoton.2013.177}{Nature Photonics {\bf 7}, 613--619}~(2013).

\bibitem{audenaert2012quantum}
Koenraad~MR Audenaert, Mil{\'a}n Mosonyi, and Frank Verstraete.
\newblock ``Quantum state discrimination bounds for finite sample size''.
\newblock \href{https://dx.doi.org/https://doi.org/10.1063/1.4768252}{Journal of Mathematical Physics{\bf 53}}~(2012).

\bibitem{sugiyama2012adaptive}
Takanori Sugiyama, Peter~S Turner, and Mio Murao.
\newblock ``Adaptive experimental design for one-qubit state estimation with finite data based on a statistical update criterion''.
\newblock \href{https://dx.doi.org/https://doi.org/10.1103/PhysRevA.85.052107}{Physical Review A {\bf 85}, 052107}~(2012).

\bibitem{sugiyama2015precision}
Takanori Sugiyama.
\newblock ``Precision-guaranteed quantum metrology''.
\newblock \href{https://dx.doi.org/https://doi.org/10.1103/PhysRevA.91.042126}{Physical Review A {\bf 91}, 042126}~(2015).

\bibitem{rubio2019quantum}
Jes{\'u}s Rubio and Jacob Dunningham.
\newblock ``Quantum metrology in the presence of limited data''.
\newblock \href{https://dx.doi.org/10.1088/1367-2630/ab098b}{New Journal of Physics {\bf 21}, 043037}~(2019).

\bibitem{meyer2023quantum}
Johannes~Jakob Meyer, Sumeet Khatri, Daniel~Stilck Fran{\c{c}}a, Jens Eisert, and Philippe Faist.
\newblock ``Quantum metrology in the finite-sample regime''~(2023).

\bibitem{helstrom1967minimum}
Carl~W Helstrom.
\newblock ``Minimum mean-squared error of estimates in quantum statistics''.
\newblock \href{https://dx.doi.org/https://doi.org/10.1016/0375-9601(67)90366-0}{Physics letters A {\bf 25}, 101--102}~(1967).

\bibitem{helstrom1968minimum}
Carl~W Helstrom.
\newblock ``The minimum variance of estimates in quantum signal detection''.
\newblock \href{https://dx.doi.org/10.1109/TIT.1968.1054108}{IEEE Transactions on information theory {\bf 14}, 234--242}~(1968).

\bibitem{helstrom1969quantum}
Carl~W Helstrom.
\newblock ``Quantum detection and estimation theory''.
\newblock \href{https://dx.doi.org/https://doi.org/10.1007/BF01007479}{Journal of Statistical Physics {\bf 1}, 231--252}~(1969).

\bibitem{helstrom1976}
Carl~W Helstrom.
\newblock ``Quantum detection and estimation theory''.
\newblock Academic press. ~(1976).
\newblock  url:~\url{https://www.sciencedirect.com/bookseries/mathematics-in-science-and-engineering/vol/123/suppl/C}.

\bibitem{petz1996monotone}
D{\'e}nes Petz.
\newblock ``Monotone metrics on matrix spaces''.
\newblock \href{https://dx.doi.org/https://doi.org/10.1016/0024-3795(94)00211-8}{Linear algebra and its applications {\bf 244}, 81--96}~(1996).

\bibitem{petzbook}
D{\'e}nes Petz.
\newblock ``Quantum information theory and quantum statistics''.
\newblock \href{https://dx.doi.org/https://doi.org/10.1007/978-3-540-74636-2}{Springer Science \& Business Media}. ~(2007).

\bibitem{hayashi-book}
Masahito Hayashi.
\newblock ``Quantum information theory''.
\newblock \href{https://dx.doi.org/https://doi.org/10.1007/978-3-662-49725-8}{Springer}. ~(2016).

\bibitem{sbd16}
Magdalena Szczykulska, Tillmann Baumgratz, and Animesh Datta.
\newblock ``Multi-parameter quantum metrology''.
\newblock \href{https://dx.doi.org/https://doi.org/10.1080/23746149.2016.1230476}{Advances in Physics: X {\bf 1}, 621--639}~(2016).

\bibitem{demkowicz2020multi}
Rafa{\l} Demkowicz-Dobrza{\'n}ski, Wojciech G{\'o}recki, and M{\u{a}}d{\u{a}}lin Gu{\c{t}}{\u{a}}.
\newblock ``Multi-parameter estimation beyond quantum fisher information''.
\newblock \href{https://dx.doi.org/10.1088/1751-8121/ab8ef3}{Journal of Physics A: Mathematical and Theoretical {\bf 53}, 363001}~(2020).

\bibitem{suzuki2020quantum}
Jun Suzuki, Yuxiang Yang, and Masahito Hayashi.
\newblock ``Quantum state estimation with nuisance parameters''.
\newblock \href{https://dx.doi.org/10.1088/1751-8121/ab8b78}{Journal of Physics A: Mathematical and Theoretical {\bf 53}, 453001}~(2020).

\bibitem{young1975}
Tzay~Y Young.
\newblock ``Asymptotically efficient approaches to quantum-mechanical parameter estimation''.
\newblock \href{https://dx.doi.org/https://doi.org/10.1016/0020-0255(75)90016-X}{Information Sciences {\bf 9}, 25--42}~(1975).

\bibitem{nagaoka1989anew}
Hiroshi Nagaoka.
\newblock ``A new approach to cram\'er-rao bounds for quantum state estimation''.
\newblock \href{https://dx.doi.org/https://doi.org/10.1142/9789812563071_0009}{IEICE Tech Report {\bf IT 89-42}, 9--14}~(1989).

\bibitem{braunstein1994statistical}
Samuel~L Braunstein and Carlton~M Caves.
\newblock ``Statistical distance and the geometry of quantum states''.
\newblock \href{https://dx.doi.org/https://doi.org/10.1103/PhysRevLett.72.3439}{Physical Review Letters {\bf 72}, 3439}~(1994).

\bibitem{hayashi2023tight}
Masahito Hayashi and Yingkai Ouyang.
\newblock ``Tight cram{\'e}r-rao type bounds for multiparameter quantum metrology through conic programming''.
\newblock \href{https://dx.doi.org/https://doi.org/10.22331/q-2023-08-29-1094}{Quantum {\bf 7}, 1094}~(2023).

\bibitem{zhang2022quanestimation}
Mao Zhang, Huai-Ming Yu, Haidong Yuan, Xiaoguang Wang, Rafa{\l} Demkowicz-Dobrza{\'n}ski, and Jing Liu.
\newblock ``Quanestimation: An open-source toolkit for quantum parameter estimation''.
\newblock \href{https://dx.doi.org/https://doi.org/10.1103/PhysRevResearch.4.043057}{Physical Review Research {\bf 4}, 043057}~(2022).

\bibitem{kimizu2024adaptive}
Masataka Kimizu, Fuyuhiko Tanaka, and Akio Fujiwara.
\newblock ``Adaptive quantum state estimation for two optical point sources''.
\newblock \href{https://dx.doi.org/https://doi.org/10.1103/PhysRevA.109.032434}{Physical Review A {\bf 109}, 032434}~(2024).

\bibitem{QestOptPOVMgit}
Jian~Chao Zhang.
\newblock ``Qestoptpovm''.
\newblock \url{https://github.com/ZHANGJianchao97/Qest}~(2023).

\bibitem{fedorovbook}
Valerii~Vadimovich Fedorov.
\newblock ``Theory of optimal experiments''.
\newblock Academic Press. ~(1972).
\newblock  url:~\url{https://lib.ugent.be/catalog/rug01:000474727}.

\bibitem{pukelsheimbook}
Friedrich Pukelsheim.
\newblock ``Optimal design of experiments''.
\newblock \href{https://dx.doi.org/https://doi.org/10.1137/1.9780898719109}{SIAM}. ~(2006).

\bibitem{fujiwara06consistency}
Akio Fujiwara.
\newblock ``Strong consistency and asymptotic efficiency for adaptive quantum estimation problems''.
\newblock \href{https://dx.doi.org/10.1088/0305-4470/39/40/014}{Journal of Physics A: Mathematical and General {\bf 39}, 12489}~(2006).

\bibitem{yamagata2011}
Koichi Yamagata.
\newblock ``Efficiency of quantum state tomography for qubits''.
\newblock \href{https://dx.doi.org/https://doi.org/10.1142/S0219749911007551}{International Journal of Quantum Information {\bf 9}, 1167--1183}~(2011).

\bibitem{suzuki2021quantum}
Jun Suzuki.
\newblock ``Quantum-state estimation problem via optimal design of experiments''.
\newblock \href{https://dx.doi.org/https://doi.org/10.1142/S0219749920400079}{International Journal of Quantum Information {\bf 19}, 2040007}~(2021).

\bibitem{d2005classical}
Giacomo~Mauro D'Ariano, Paoloplacido~Lo Presti, and Paolo Perinotti.
\newblock ``Classical randomness in quantum measurements''.
\newblock \href{https://dx.doi.org/10.1088/0305-4470/38/26/010}{Journal of Physics A: Mathematical and General {\bf 38}, 5979}~(2005).

\bibitem{hayashi97}
M~Hayashi.
\newblock ``A linear programming approach to attainable cramer-rao type bound''.
\newblock In Osamu Hirota, Alexander~S Holevo, and Carlton~M Caves, editors, Quantum Communication, Computing, and Measurement.
\newblock \href{https://dx.doi.org/https://doi.org/10.1007/978-1-4615-5923-8_11}{Pages 150--–161}.
\newblock Plenum, New York~(1997).

\bibitem{GM00}
Richard~D Gill and Serge Massar.
\newblock ``State estimation for large ensembles''.
\newblock \href{https://dx.doi.org/https://doi.org/10.1103/PhysRevA.61.042312}{Physical Review A {\bf 61}, 042312}~(2000).

\bibitem{holevobook}
Alexander~S Holevo.
\newblock ``Probabilistic and statistical aspects of quantum theory''.
\newblock \href{https://dx.doi.org/https://doi.org/10.1007/978-88-7642-378-9}{Edizioni della Normale}. ~(2011).

\bibitem{nagaoka91}
Hiroshi Nagaoka.
\newblock ``A generalization of the simultaneous diagonalization of hermitian matrices and its relation to quantum estimation theory''.
\newblock In Masahito Hayashi, editor, Asymptotic Theory Of Quantum Statistical Inference: Selected Papers.
\newblock \href{https://dx.doi.org/https://doi.org/10.1142/9789812563071_0012}{Pages 133--149}.
\newblock World Scientific~(2005).

\bibitem{conlon2021efficient}
Lorc{\'a}n~O Conlon, Jun Suzuki, Ping~Koy Lam, and Syed~M Assad.
\newblock ``Efficient computation of the nagaoka--hayashi bound for multiparameter estimation with separable measurements''.
\newblock \href{https://dx.doi.org/https://doi.org/10.1038/s41534-021-00414-1}{npj Quantum Information {\bf 7}, 110}~(2021).

\bibitem{fn99}
Akio Fujiwara and Hiroshi Nagaoka.
\newblock ``An estimation theoretical characterization of coherent states''.
\newblock \href{https://dx.doi.org/https://doi.org/10.1063/1.532962}{Journal of Mathematical Physics {\bf 40}, 4227--4239}~(1999).

\bibitem{conlon2022gap}
Lorc{\'a}n~O Conlon, Jun Suzuki, Ping~Koy Lam, and Syed~M Assad.
\newblock ``The gap persistence theorem for quantum multiparameter estimation''~(2022).

\bibitem{boyd2004convex}
Stephen~P Boyd and Lieven Vandenberghe.
\newblock ``Convex optimization''.
\newblock \href{https://dx.doi.org/https://doi.org/10.1017/CBO9780511804441}{Cambridge university press}. ~(2004).

\bibitem{somim}
Kean~Loon Lee, Jiangwei Shang, Wee~Kang Chua, Shiang~Yong Looi, and Berthold-Georg Englert.
\newblock ``Somim: An open-source program code for the numerical search for optimal measurements by an iterative method''~(2008).

\bibitem{scs}
Brendan O'Donoghue, Eric Chu, Neal Parikh, and Stephen Boyd.
\newblock ``Conic optimization via operator splitting and homogeneous self-dual embedding''.
\newblock \href{https://dx.doi.org/https://doi.org/10.1007/s10957-016-0892-3}{Journal of Optimization Theory and Applications {\bf 169}, 1042--1068}~(2016).

\bibitem{rehacek2004minimal}
Jaroslav {\v{R}}eh{\'a}{\v{c}}ek, Berthold-Georg Englert, and Dagomir Kaszlikowski.
\newblock ``Minimal qubit tomography''.
\newblock \href{https://dx.doi.org/https://doi.org/10.1103/PhysRevA.70.052321}{Physical Review A {\bf 70}, 052321}~(2004).

\bibitem{YL73}
Horace Yuen and Melvin Lax.
\newblock ``Multiple-parameter quantum estimation and measurement of nonselfadjoint observables''.
\newblock \href{https://dx.doi.org/https://doi.org/10.1109/TIT.1973.1055103}{IEEE Transactions on Information Theory {\bf 19}, 740--750}~(1973).

\bibitem{KM}
Keiji Matsumoto.
\newblock ``A new approach to the cram{\'e}r-rao-type bound of the pure-state model''.
\newblock \href{https://dx.doi.org/10.1088/0305-4470/35/13/307}{Journal of Physics A: Mathematical and General {\bf 35}, 3111}~(2002).

\end{thebibliography}

\end{document}